\documentclass{jpp}
\usepackage{graphicx}
\usepackage{epstopdf, epsfig}

\usepackage{color}

\newcommand{\bgeqa}{\begin{eqnarray}}
\newcommand{\edeqa}{\end{eqnarray}}
\newcommand{\refeq}[1]{Eq.~(\ref{#1})}

\definecolor{update}{rgb}{1,0,0} 
\definecolor{update}{rgb}{0,0,0} 

\shorttitle{Generalized fluid theory including non-Maxwellian kinetic effects}
\shortauthor{O. Izacard}

\title{Generalized fluid theory including non-Maxwellian kinetic effects}

\author{Olivier Izacard\aff{1}
  \corresp{\email{izacard@llnl.gov}}}

\affiliation{\aff{1}Lawrence Livermore National Laboratory, 7000 East Avenue, L-637, Livermore, California 94550, USA}

\begin{document}

\maketitle
{\centerline{14 July 2016}}

\begin{abstract}
The results obtained by the plasma physics community for the validation and the prediction of turbulence and transport in magnetized plasma come mainly from the use of very CPU-consuming particle-in-cell or (gyro)kinetic codes which naturally include non-Maxwellian kinetic effects. To date, fluid codes are not considered to be relevant for the description of these kinetic effects. Here, after revisiting the limitations of the current fluid theory developed  {\color{update}in the 19th century}, we generalize the fluid theory including kinetic effects such as non-Maxwellian super-thermal tails with as few fluid equations as possible. The collisionless and collisional fluid closures from the nonlinear Landau Fokker-Planck collision operator are shown for an arbitrary collisionality. Indeed, the first fluid models associated with two examples of collisionless fluid closures are obtained by assuming an analytic non-Maxwellian distribution function (e.g., the INMDF {\color{update}[O. Izacard, \textit{Phys. Plasmas} \textbf{23}, 082504 (2016) Kinetic corrections from analytic non-Maxwellian distribution functions in magnetized plasmas]}). One of the main differences with the literature is our analytic representation of the distribution function in the velocity phase space with as few hidden variables as possible thanks to the use of non-orthogonal basis sets. These new non-Maxwellian fluid equations could initiate the next generation of fluid codes including kinetic effects and can be expanded to other scientific disciplines such as astrophysics, condensed matter, or hydrodynamics. {\color{update}As a validation test, we perform a numerical simulation based on a minimal reduced INMDF fluid model. The result of this test is the discovery of the origin of particle and heat diffusion. The diffusion is due to the competition between a growing INMDF on short time scales due to spatial gradients and the thermalization on longer time scales.} The results shown here could draw the breaking of some unsolved understandings of the turbulence.
\end{abstract}

\section{Introduction}
\label{ToC:intro}
The description of turbulence using the fluid theory has been developed as early as in the 19th century~\citep{Pomeau} whereas the kinetic theory has followed a slightly different path~\citep{Brush_2003}. One of the most important unsolved problems in plasma physics is the unification between kinetic and fluid descriptions. The fluid description is advantageous because of its simplicity, its relatively light numerical demands and its use in reduced modeling. However, as explained here, the existing fluid theory is always related to a steady-state Maxwellian distribution function (MDF). In contrast, the kinetic description has the advantage of taking into account non-Maxwellian (NM) kinetic effects. As a result, after more than $30$ years of code developing efforts made by the community, highly CPU-consuming particles-in-cell~\citep{Dawson_1968,
Birdsall_1968,
Fonseca_2002,
GTS,
GTC,
GEM,
GT3D,
ORB5_2007} and (gyro)kinetic~\citep{GS2_1995,
GS2_2000,
GENE,
GYRO,
GKV,
GYSELA,
ELMFIRE,
COGENT} codes dominate the field. A large part of the current work of the community aims at expanding the kinetic short time scale results into longer time scales compatible with confinement time~\citep{TGLF,Parra_2008,Dorf_2016}. Indeed, the anomalous transport coefficients are given by a short time scale code which include kinetic effects such as {\color{update}GYRO~\citep{GYRO}}, and are used inside a longer time scale fluid code such as {\color{update}TGLF~\citep{TGLF}} for the simulation of the long time scale dynamics. The transport coefficients (i.e., the particle density diffusion and the heat conductivity) can be computed as function of one or two spatial dimensions in order to describe different transport at different position in the tokamaks (e.g., core or scrape-off-layer). A similar technique using profiles of the transport coefficients is used in fluid simulations in order to match experimental profiles of measured quantities (e.g., density and temperature profiles at the outer midplane of tokamaks) during the validation~\citep{Chankin_2007_NF_47a,
Chankin_2007_NF_47b,
Groth_2013_NF_53} of the state-of-the-art edge fluid simulations against experimental measurements. 
For all the reasons mentioned above, it is therefore of great interest to develop undiscovered links between kinetic and fluid descriptions that have advantages of both: it would need relatively small numerical resources, it would be relevant to short and long time scales, and it would include NM kinetic effects. \\
Moreover, even for most (gyro)kinetic and drift-kinetic codes, the background steady state distribution function is assumed to be a MDF and the NM deviations are described by the fluctuations which by construction need to have no temporal mean value. In other words, it is not self-consistent to describe NM steady states by assuming a MDF for the non-fluctuating part. Experimental observations of discrepancies due to kinetic effects~\citep{Mazon_2016_Nature,
Luna_2003_RSI,
Beausang_2011_RSI,
Taylor_1996_PPCF,
Fidone_1996_PoP,
Bitter_2003_PRL,
Bartiromo_1986_NF} and predictions of kinetic corrections from analytic NMDFs~\citep{Izacard_2016_FischSymposium,Izacard_2016_Paper1} can be significant, even for a small population of non-thermalized particles, so it is indispensable to take into account NM steady states. As shown by~\citet{Izacard_2016_FischSymposium,Izacard_2016_Paper1}, the advantage of assuming an analytic NMDF (instead of a MDF) is such that we can keep some kinetic effects and analytically compute the velocity phase space integrals for the secondary electron emission, the Langmuir probe characteristic curve and the entropy. One of the used analytic NMDF introduced in that work is the interpreted NMDF (called INMDF in Ref.~\citep{Izacard_2016_Paper1} and generalized in other INMDFs by using a non-orthogonal basis set) which describes a displacement of a population of particles in the velocity phase space. It has been shown that this simple formulation allows a better understanding of the electron temperature discrepancy (up to $20\%$) between the Thomson scattering and the electron cyclotron emission measurements in JET and TFTR tokamaks because it fits very well~\citep{Izacard_2016_Paper1} the numerical model NMDF computed in Refs.~\citep{Luna_2003_RSI,Beausang_2011_RSI}. This fact is the proof that INMDFs exist in experiments. Moreover, INMDFs can also be easily fitted to the NMDFs numerically computed from 3D Fokker-Planck or particles-in-cell codes~\citep{Izacard_2016_FischSymposium}. Given these results it makes sense to perform another step and to think about the fluid reduction from the assumed INMDFs. The fluid models developed here are associated with the first INMDF but can be generalized by the reader as needed. As it is explained here, it has never been possible to rigorously and efficiently include this kind of kinetic effects into fluid models since some assumptions and choices used in the literature were not optimal. \\
Moreover, because this fluid reduction from NMDFs can be performed, we could, in the future, be able to solve some problems such as the radiation shortfall between the current state-of-the-art fluid simulations and the experimental measurements, particularly in detached plasmas. Indeed, some recent work~\citep{Chankin_2007_NF_47a,
Chankin_2007_NF_47b,
Groth_2013_NF_53} argue that the radiation shortfall exists even by including cross magnetic field drifts, and it seems to be due to kinetic effects (i.e., the presence of NMDFs). Then, instead of developing gyrokinetic codes for the edge, it will be greatly advantageous to develop a generalization of the fluid theory in order to include kinetic effects in very simple fluid models. Indeed, one of the main advantages of using fluid equations is the simplicity of the models in comparison to the fact that even after 30 years of research some researchers do not agree about the validity of the gyrokinetic equations~\citep{Parra_2008,Lee_2009,Parra2,Lee2}. \\
Sec.~\ref{ToC:limitations} contains criticisms of the current state of the fluid theory. This will help the reader to understand some reasons of the missing development of the efficient generalized fluid theory for NMDFs. Sec.~\ref{ToC:NewFluid} details the first non-Maxwellian fluid models consistent with a localized super-thermal population of particles with two different collisionless fluid closures and the description of the collisional fluid closure from the full nonlinear Landau Fokker-Planck collision operator~\citep{Landau_1946,Rosenbluth}. Finally, Sec.~\ref{ToC:CCL} is dedicated to the discussion of the generalization of the fluid theory and the listing of perspectives motivated by this work.

\section{Current state of the fluid theory}
\label{ToC:limitations}
This section focuses on the limitations of the current fluid theory. Before giving the solution of fluid equations associated with the INMDF, we need to understand the limitations of the current fluid theory in order to see which assumptions can be modified. A summary of the current understanding of the fluid theory is well represented by the first sentences of Ref.~\citep{Plunk_2010_JFM_664}
\begin{quote}
``A fluid is conventionally  described  by  macroscopic  state  variables  such  as  bulk flow velocity, density, pressure, etc., which vary over three-dimensional space. This description is appropriate when collisions between the constituent particles establish local thermodynamic equilibrium (Maxwellian velocity distribution) more rapidly than any dynamical processes that can disturb this equilibrium. When this condition is not met, a kinetic description is needed to capture the evolution of a distribution function in six-dimensional phase space (positions and velocities).''
\end{quote}
A large number of similar citations can be found in the literature. We describe in this section how the current fluid theory has led to misunderstandings that are commonly accepted. One of the main reasons of these misunderstandings is the word ``etc.'' in the citation above. By avoiding the efficient description of other fluid quantities, we cannot easily describe non-Maxwellian steady-state distribution functions. Indeed, before generalizing the fluid theory associated with NMDFs we have to clarify the construction of NMDFs using a small number of fluid quantities. It is shown in Sec.~\ref{ToC:NewFluid} that it is possible to obtain a small number of new fluid equations relevant to NMDFs at finite collisionality (i.e., when the characteristic time between collisions is not negligible in front of the characteristic time of all dynamical processes).
In this section, some details introducing the results shown in Sec.~\ref{ToC:NewFluid} are disseminated in order to better point out the limitations of the current fluid theory.
\subsection{Limitations of the current fluid theory}
\begin{figure}
\centerline{
(a)\label{fig.Mk.Math.Hermite}
\includegraphics[height=40mm]{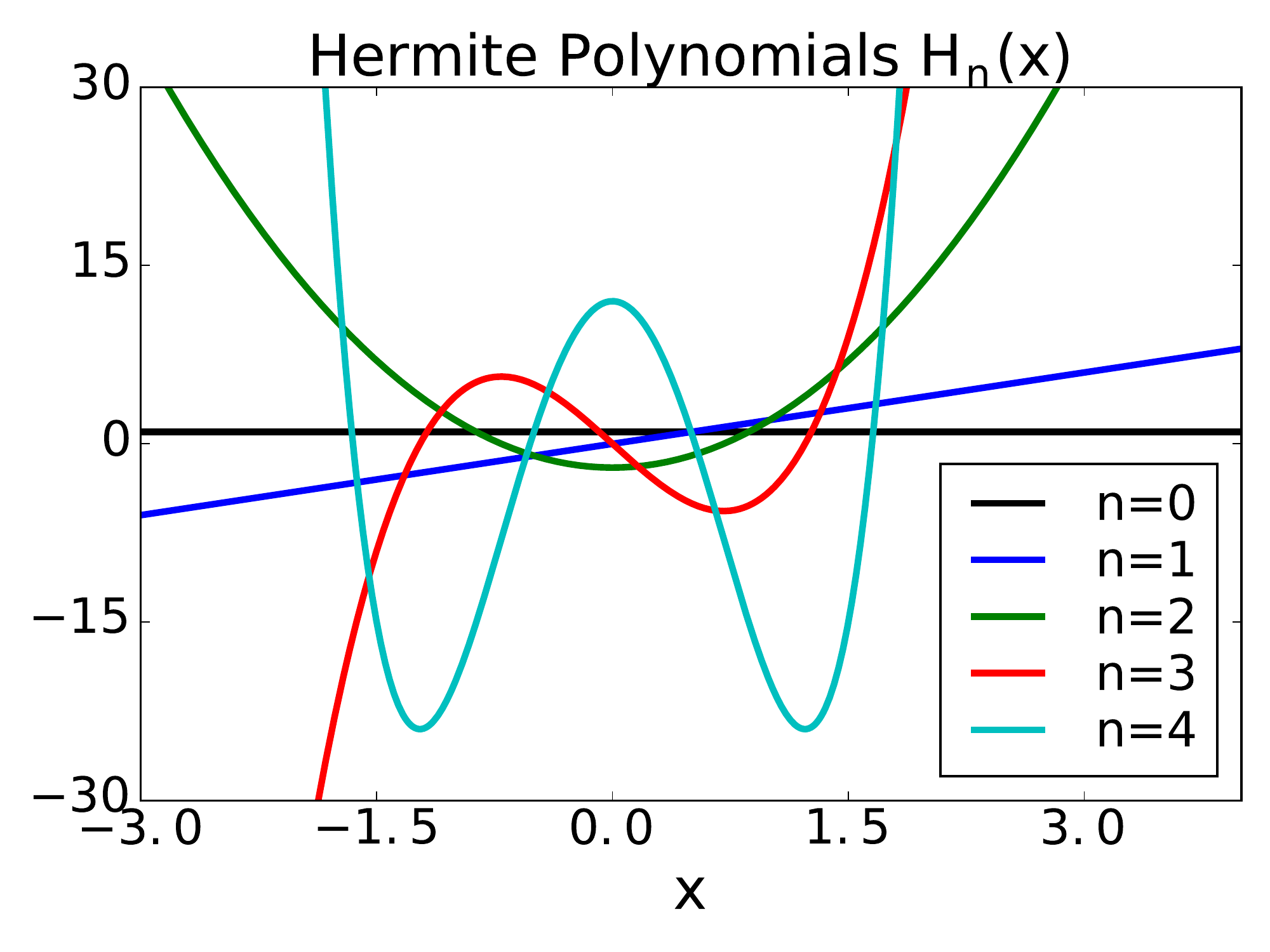}
(b)\label{fig.Mk.Math.Laguerre}
\includegraphics[height=40mm]{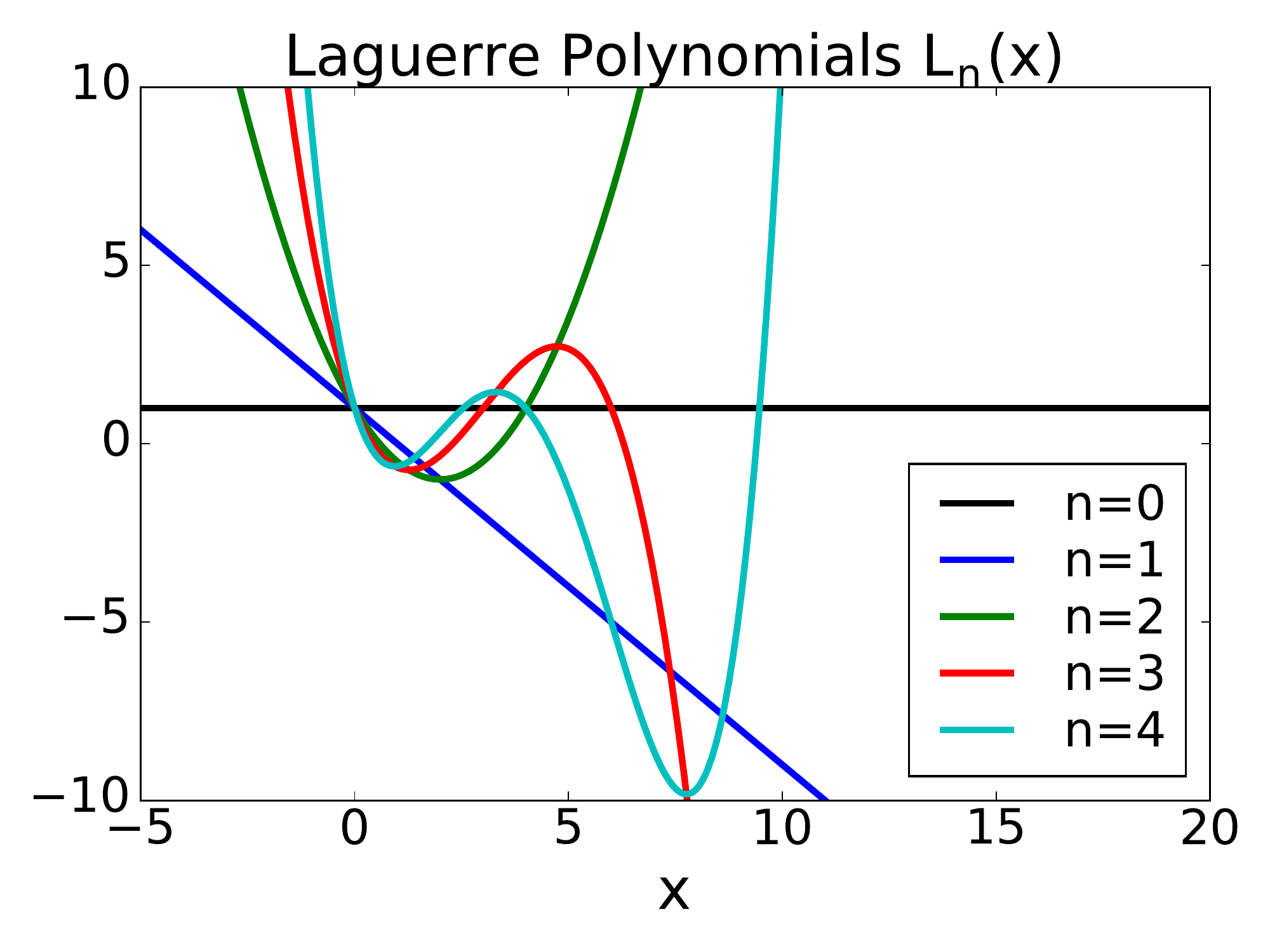}
}
\centerline{
(c)\label{fig.Mk.Math.Fourier1}
\includegraphics[height=40mm]{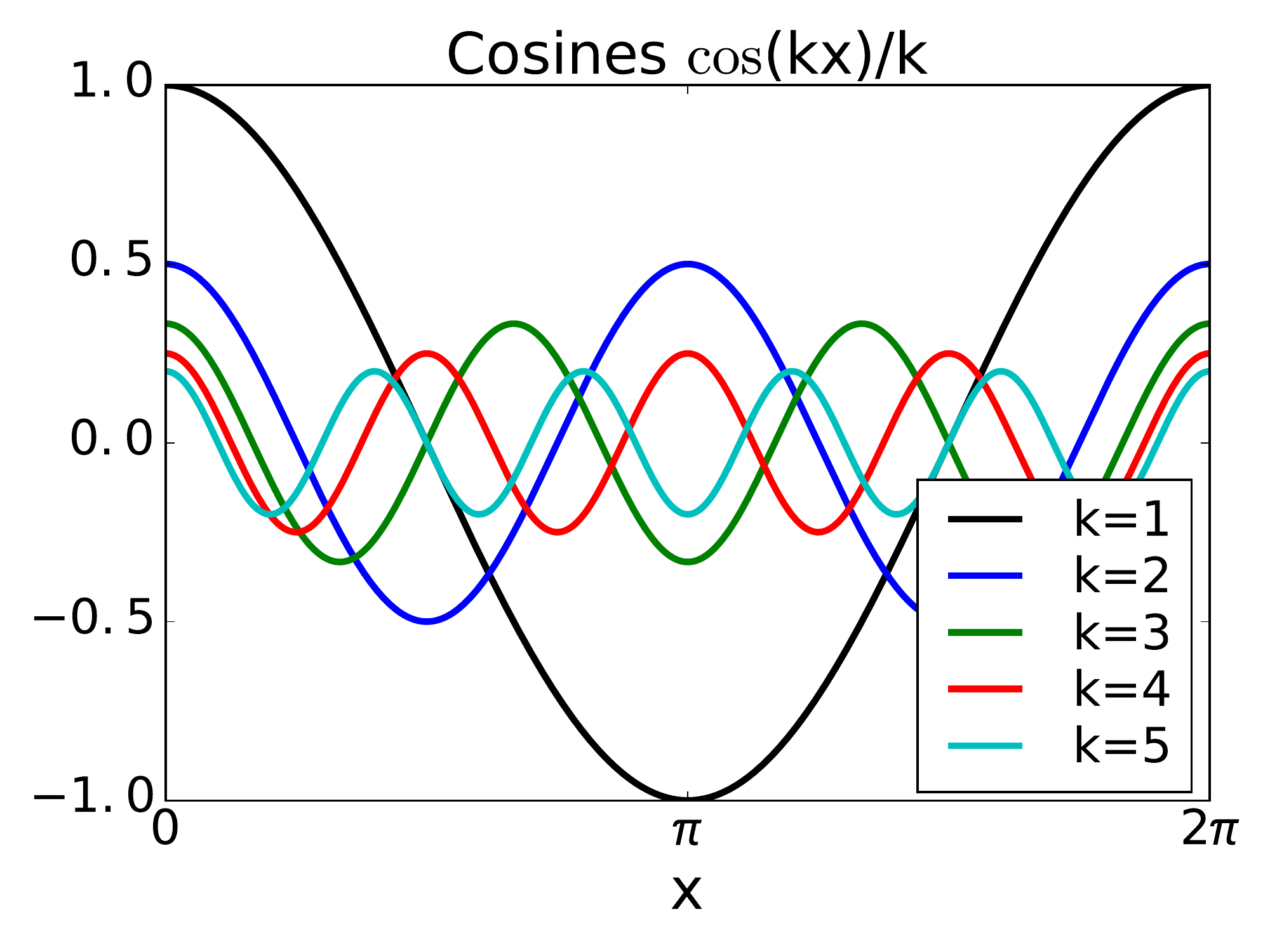}
(d)\label{fig.Mk.Math.Fourier2}
\includegraphics[height=40mm]{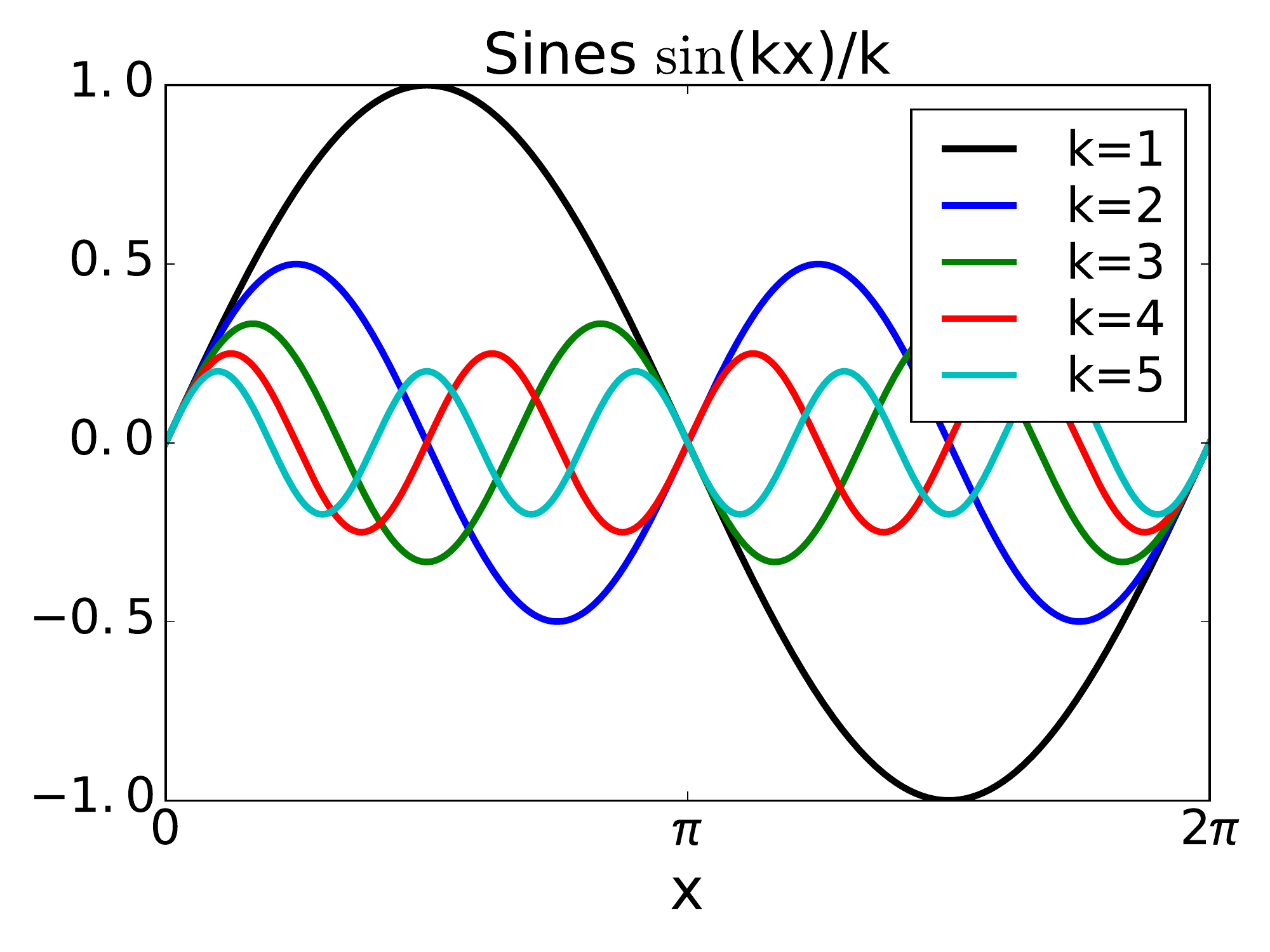}
}
\centerline{
(e)\label{fig.Mk.Math.Bessel}
\includegraphics[height=40mm]{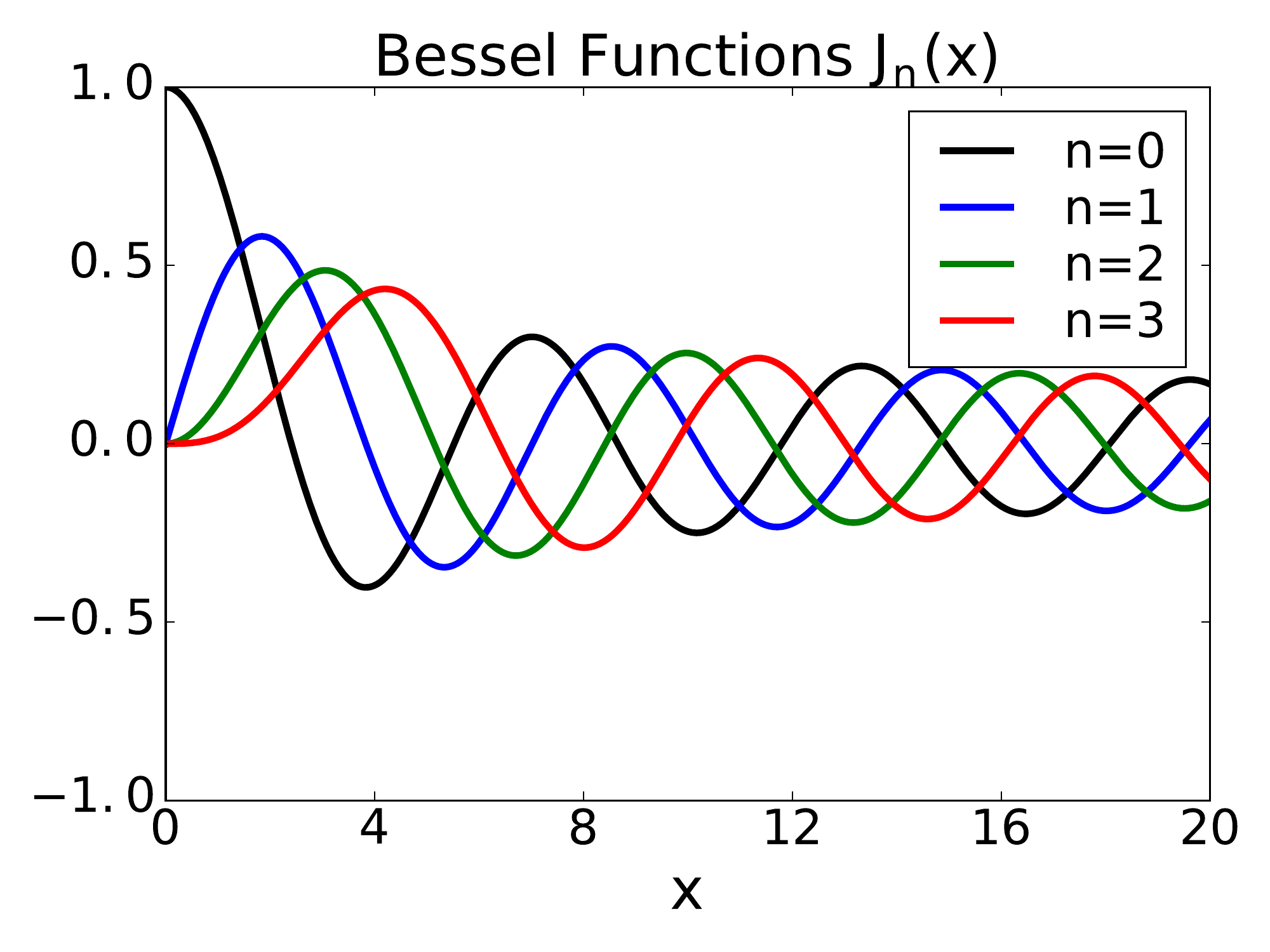}
(f)\label{fig.Mk.Math.Legendre}
\includegraphics[height=40mm]{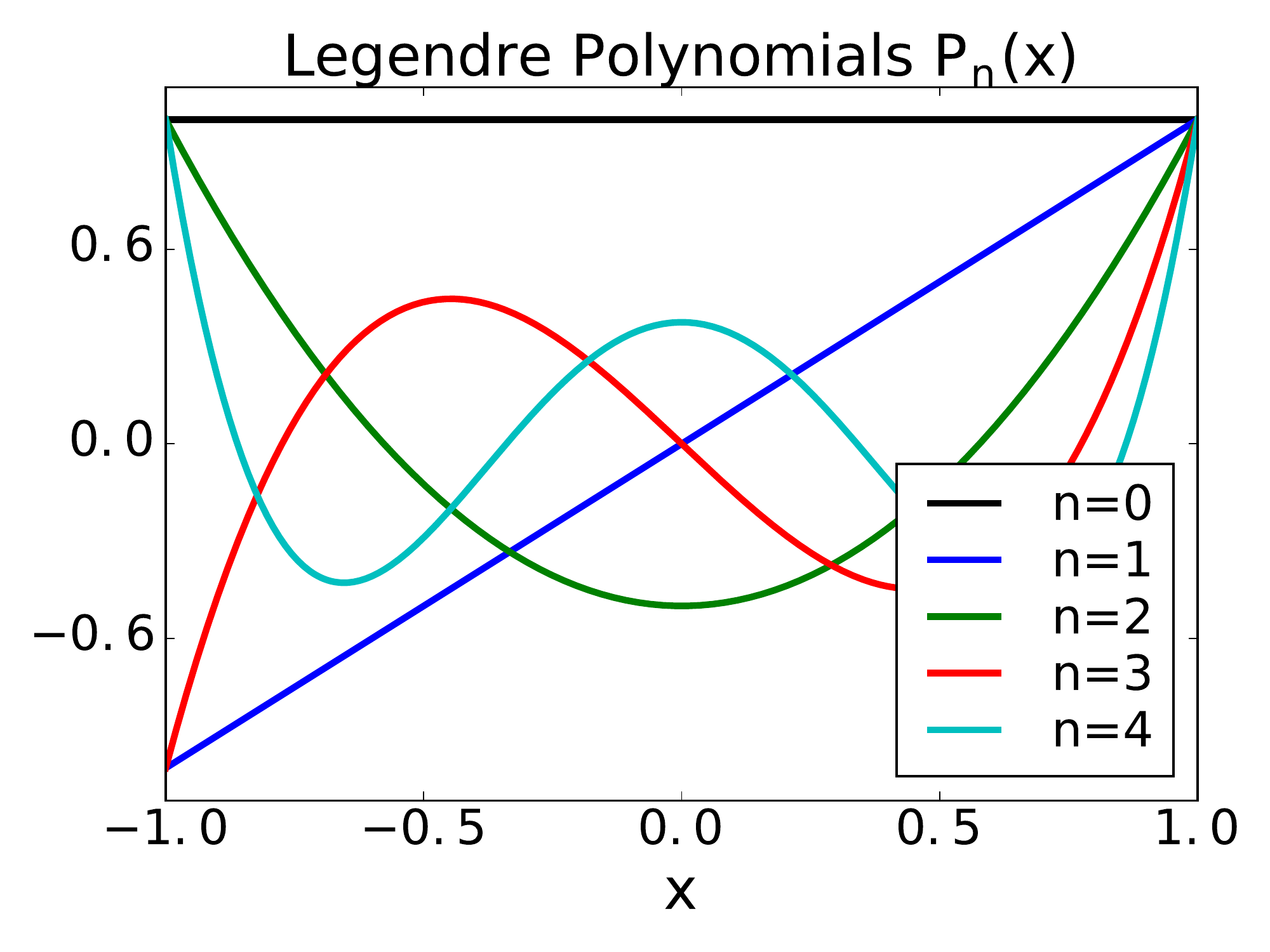}
}
\caption{Orthogonal basis functions commonly used to represent NMDFs. The limit at infinity of Hermite (a), Laguerre (b) and Fourier (c)-(d) basis functions are not $0$. The Bessel basis functions (e) are quasi-periodic and the Legendre basis functions (f) are relevant only for discretized distribution functions. None of them are efficient (e.g., a MDF can be describe with at least a few hundreds of terms as function of the accuracy).}
\label{fig.Mk.Math}
\end{figure}

{\color{update}The fluid theory was developed for ideal fluid (i.e., related to a MDF) by Euler in the 18th century, and more significantly for non-ideal flows in the 19th century with the well known works of Navier, Cauchy, Poisson, Stokes, Reynolds, and others who found fluid equations describing transport and turbulence. In parallel, the kinetic theory of gas was developed and allowed for a deeper description and understandings of transport and turbulence. However, the fluid theory has not been able to recover some results of the kinetic theory yet. Indeed, the} first limitation which has been recurrent in the last century is the use of non-adapted basis functions for the representation of the distribution function. One of the first noticeable works linking non-Maxwellian deviations of the distribution function and the fluid moments was published by~\citet{Grad_1949_CPAM_2,Grad_1963_PoF_6}. In these references, Grad used the Hermite polynomials and obtained the well known $13$-moments fluid model. Since this result was obtained, all analytic representations of the distribution function have used one of the following orthogonal basis sets: Hermite, Laguerre or Legendre polynomials, the Bessel functions or the Fourier series (see Fig.~\ref{fig.Mk.Math}).

These representations are not efficient for the description of non-Maxwellian steady-state distribution functions or even for the fluctuation part. Indeed, for the Hermite and Laguerre polynomials as well as for the Fourier series, their limit at infinity is not $0$ so these representations are not adapted for localized modifications of the distribution function. Moreover, due to the fact that these basis functions are not compact, a very large number of terms are required to describe a compact deviation of the distribution function. The multiplication of these basis functions by a background Maxwellian (e.g., \citep{Grad_1949_CPAM_2}) is still not efficient enough even if the distribution function becomes compact because the departure of the distribution function from the Maxwellian (i.e., $f/f_0$ where $f$ is the NMDF and $f_0$ the MDF) is also not well described by all of these orthogonal basis functions~\citep{Izacard_2016_FischSymposium}. For the Bessel functions and the Legendre polynomials, their limit goes to $0$ but the quasi-periodicity of the Bessel functions and the non-efficiency of the Legendre polynomials are additional reasons of the failure toward the efficient description of NMDFs. Indeed, the Legendre polynomials are adapted for a discretized distribution function which is not efficient for the fluid reduction due to the large number of terms (hence, the large number of fluid equations). Furthermore, even for the description of a MDF, one needs a few hundreds or thousands of terms with a discretized distribution function instead of only $3$ parameters with the continuous exponential function. At the end, the fluid reduction can be efficient only by using continuous, non-quasi-periodic, and compact basis functions which is the case for the MDF, after all. An important related limitation found in Ref.~\citep{Hirvijoki:15} %
 is the use of a sum of shifted MDFs (called in the literature the radial Gaussian basis functions, RGBF)~\footnote{We note the link between RGBF and the bi-modal distribution function (i.e., the sum of two Maxwellians) which has been used to approximate NMDFs for the correction of the Langmuir probe interpretation, omitting the self-consistent interaction (i.e., collision) between the two MDFs.}. This use is a limitation because for an arbitrary collisionality, a MDF is the steady-state solution of an isolated system (i.e., when the self-collisional time is much smaller than the characteristic time of exchange of energy with external sources). Then, it is rarely self-consistent to superpose at least two MDFs corresponding respectively to the bulk plasma and to the super-thermal population. Both reach their thermodynamic equilibrium by neglecting all interactions (i.e., collisions) with other species or injected particle sources. However, the presence of the super-thermal population is due to non-negligible collisions with a source of energy (e.g., NBI or $\alpha$-particles). The last sentence can enter in conflict with the fact that both species are at their thermodynamic equilibrium, which is reached only when we neglect the collision with other species. In summary, the sum of MDFs and more generally the RGBF are valid only at the specific regime of collisions
\bgeqa
\label{eq.nu}
\displaystyle
\frac{ \nu_{\rm a-a} }{ \nu_{\rm a-b} } \rightarrow \infty,
\edeqa
where $a \in \{{\rm th},{\rm f}\}$, $b \in \{{\rm th},{\rm f},{\rm s}\}$ and $a \neq b$ (${\rm th}$, ${\rm f}$ and ${\rm s}$ represent respectively the thermal, fast and source populations). In other words, $\nu_{\rm a-a}$ represents the self-collisionality of the thermalized or the fast population, and $\nu_{\rm a-b}$ represents the collisionality of the interaction between thermalized, fast particles or sources. This regime is too constraining for many experiments where a finite ratio needs to be taken into account. \\
Finally, one of the arguments found in the literature for the use of the representations mentioned above is that their basis functions are orthogonal. However, this argument is only needed for the conveniences of analytic developments for a very large number of terms and is not physically motivated. 
We use here the generalized formula of the interpreted NMDF (called INMDF) introduced by~\citet{Izacard_2016_FischSymposium,Izacard_2016_Paper1}
\bgeqa
\label{eq.fnM.Generalization}
\displaystyle
f &=& \sum_{k=0}^{N} \frac{ a_k \left( {\rm v} - b_k \right)^{n_k} }{\left(2\pi d_k^{m_k}\right)^{1/2}}  \exp\left( -\frac { \left( {\rm v}-c_k\right)^2 }{ 2e_k }  \right),
\edeqa
with $m_k = 2 {\rm E}[(n_k+1)/2]+1$, $n_k \in \mathbb{N}$ where ${\rm E}[x]$ is the floor function and $a_k$, $b_k$, $c_k$, $d_k$, $e_k$ are fluid coefficients for all integer $k$ ($N$ is the maximum number of terms), {\color{update}and ${\rm v}$ is the velocity phase space coordinate}. The set of coefficients $\{ a_k, b_k, c_k, d_k, e_k \}$ are the hidden variables. 
\begin{figure}
\centerline{
\includegraphics[height=65mm]{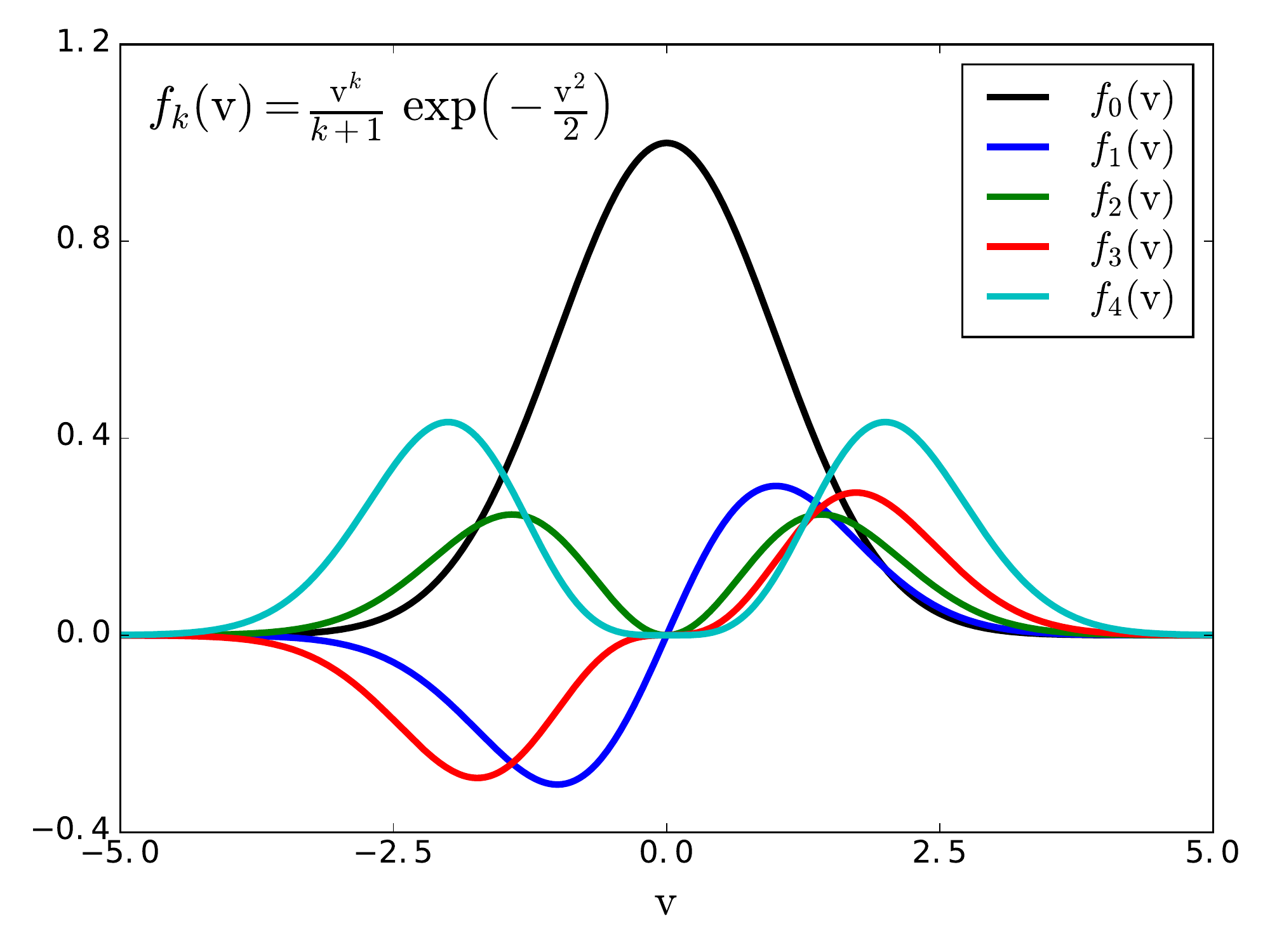}
}
\caption{Example of basis functions relevant to INMDFs. These basis functions efficiently describe localized modifications of NMDFs and parameters can be added ``\`{a} la Maxwellian'' to allow robust representations.}
\label{fig.Mk.BasisINMDF}
\end{figure}
The solution shown in Fig.~\ref{fig.Mk.BasisINMDF} solves the problem of the non-adapted {\color{update}orthogonal} basis functions used in the literature. This figure represents the first terms of possible basis functions that can efficiently describe NMDFs, especially with a super-thermal tail. We remark that these basis functions are not orthogonal in comparison to the previously mentioned ones. Moreover, this set of basis functions is so general that it can reproduce a sum of MDFs as well as the Hermite polynomials~\citep{Izacard_2013_Emails_Candy}. The use of the first basis function $f_0({\rm v})$ allows us to describe MDFs with $3$ hidden variables: the density $n$, the fluid velocity $v$ and the temperature $T$. Each of these $3$ parameters has a well known physical interpretation. Following this methodology, the INMDF (see Refs.~\citep{Izacard_2013_GA_Talk,Izacard_2016_FischSymposium,Izacard_2016_Paper1} and \refeq{eq.fI}) is the first example of a NMDF representing a super-thermal tail that uses the second basis function $f_1({\rm v})$ associated with $3$ additional hidden variables: the kinetic flux $\Gamma$, the central flow $c$ and the width of the heat spread $W$. These new parameters have been physically interpreted in Ref.~\citep{Izacard_2016_Paper1} as a displacement of a population of particles in the velocity phase space due to an external source of energy. 
\begin{figure}
\centerline{
(a)\label{fig.nM.ErrorHermite-vs-Izacardian.a}
\includegraphics[height=47mm]{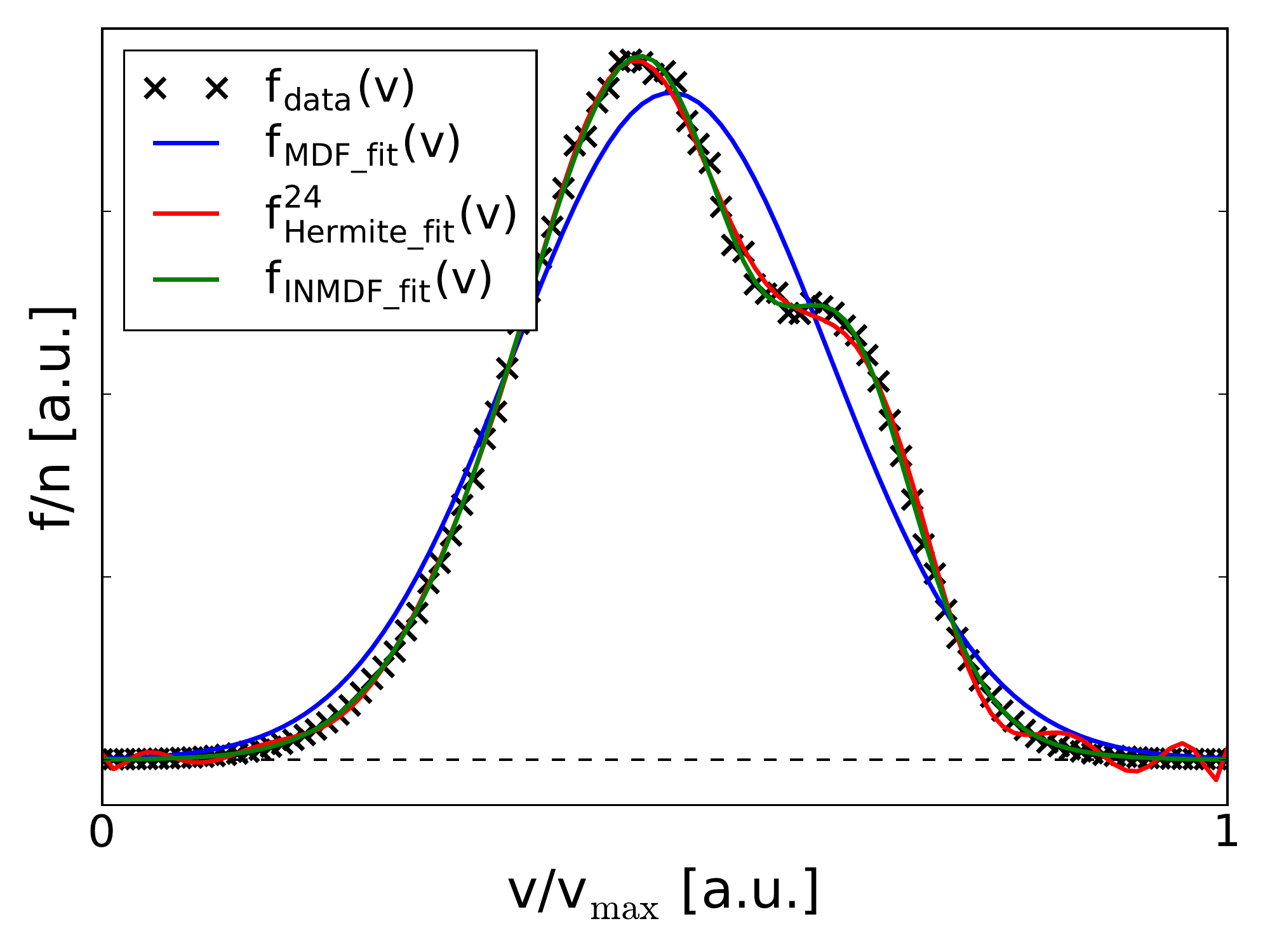}
(b)\label{fig.nM.ErrorHermite-vs-Izacardian.b}
\includegraphics[height=47mm]{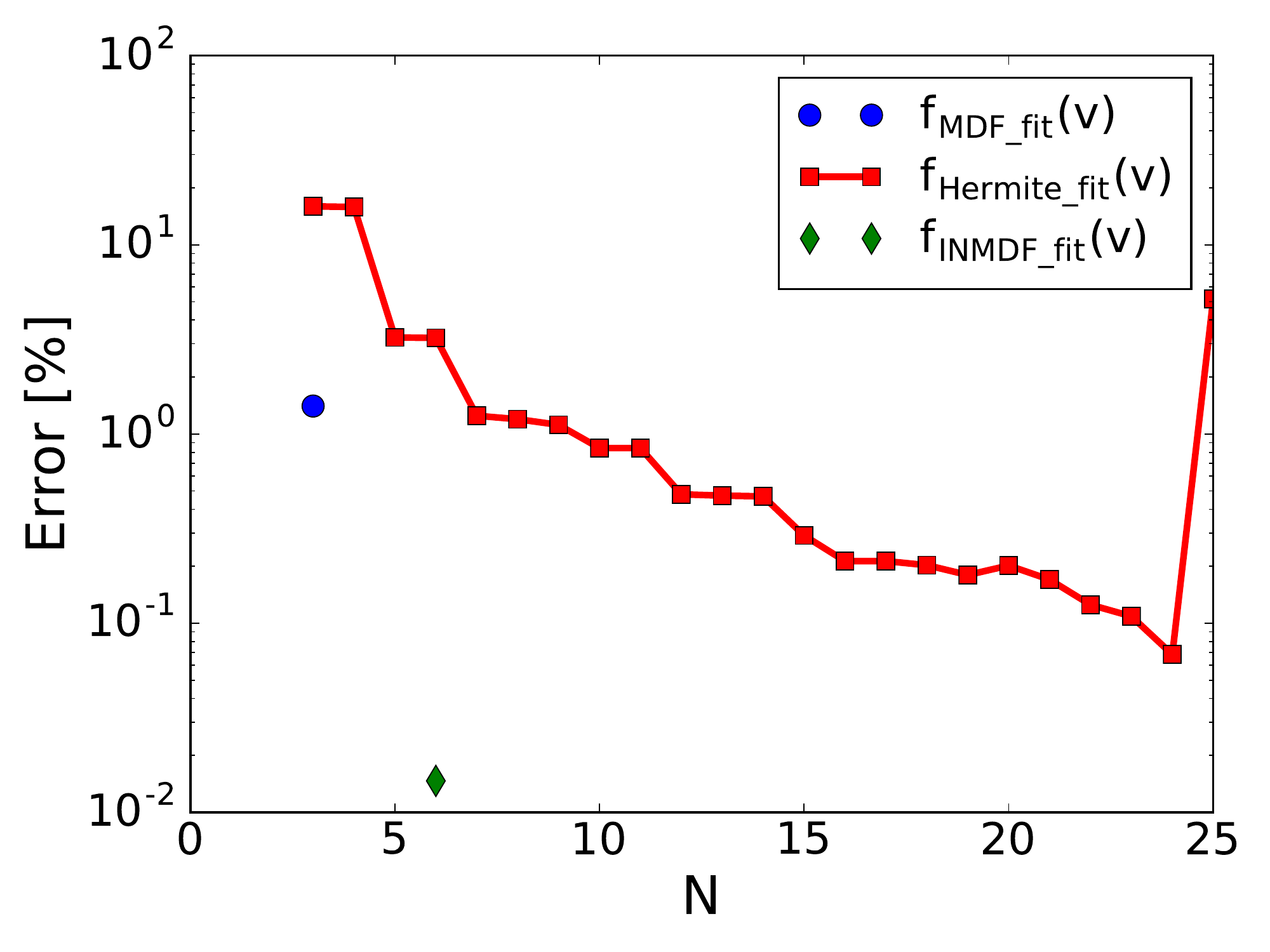}
}
\caption{Fig.~(a) represents the comparison of the fitting of artificial data (black crosses) with a Maxwellian (blue curve), a Hermite polynomials of order $N=24$ (red curve) and an INMDF (green curve). The artificial data are constructed from an INMDF and $5\%$ of noise. Fig.~(b) shows the relative error in percent between the fitted distribution function and the artificial data as function of the number of terms $N$ used by the representation of the distribution function.}
\label{fig.nM.ErrorHermite-vs-INMDF}
\end{figure}
The efficiency of the INMDF to describe a localized super-thermal tail is shown in Fig.~\ref{fig.nM.ErrorHermite-vs-INMDF} which compares different orders of the Hermite polynomials and the INMDF to fit artificial data. The artificial data are created from an INMDF which includes $4\%$ of super-thermal particles and $5\%$ of noise. The crosses in Fig.~\ref{fig.nM.ErrorHermite-vs-INMDF}.(a) are a plot of the artificial data. The blue, red and green curves correspond respectively to the least-squares fitting of the artificial data by a MDF, an INMDF, or a Hermite polynomial of order $N=24$. We remark that due to the limited number of velocity grid points (i.e., $101$ points), the best fit from a Hermite polynomial is obtained with $24$ terms (see Fig.~\ref{fig.nM.ErrorHermite-vs-INMDF}.(b)) and if we want to reduce the least-squares errors we have to increase the velocity phase-space resolution. The fitted MDF (blue curve) is defined by the three first moments of the artificial data. Fig.~\ref{fig.nM.ErrorHermite-vs-INMDF}.(b) represents the relative error in percent between the artificial data and the fitted curves as function of the number of used parameters. The Maxwellian uses $N=3$ parameters, the INMDF uses $N=6$, and the Hermite polynomial gives a minimal error (for the artificial data shown in Fig.~\ref{fig.nM.ErrorHermite-vs-INMDF}.(a)) when $24$ coefficients are kept.   The error of the Hermite polynomial is due to the fluctuations at velocities far from the fluid velocity $v$ (i.e., far away from ${\rm v}/{\rm v}_{max} \sim 0.5$) as well as the mismatch of the distribution function close to the central flow $c$ (${\rm v}/{\rm v}_{max} \sim 0.6$). In the Hermite representation, a large number of terms are required in order to reduce the divergence at infinity induced from the first added term. In other words, the Hermite polynomial is not efficient for the description of local departures of a MDF, in the same way than a polynomial will not be efficient in general for the representation of a cosine. The INMDF is clearly much more accurate and efficient in reproducing super-thermal particles with a minimum number of parameters (i.e., hidden variables) even when the super-thermal tail represents more than $4\%$ of particles as shown in Fig.~\ref{fig.nM.ErrorHermite-vs-INMDF}.(a). \\
Summarizing the first limitation given here, the choice of the basis function used for the description of NMDFs is crucial since the number of hidden variables (i.e., the fluid parameters) is completely related to this choice. The INMDFs generalized by \refeq{eq.fnM.Generalization} seems to be one of the best way to describe localized deviations from the MDF, at least in presence of super-thermal particles. The reader may develop other non-orthogonal basis sets in order to describe other phenomena. \\

A second limitation found in the literature is the acceptation of misunderstandings about the fluid theory and the fluid closure. Here is a list of some of these misunderstandings accepted by the majority of the community.\\
 \hspace{5mm}\textbullet\ F-(i) Many researchers~\footnote{Discussions at the \textit{Plasma-Material Interactions Community Workshop}, Princeton, New Jersey USA. May 4–6 2015.} agree about the fact that the fluid codes are useless with respect to (gyro)kinetic codes for the description of many phenomena in plasmas since they cannot capture kinetic effects. We must agree that, until now, no rigorous and efficient developments of fluid equations catching kinetic effects, even in the collisionless limit, have been published. Indeed, this work is the proof that the fluid theory can efficiently describe kinetic effects. The efficiency of the fluid reduction is obviously dependent on the choice of the basis functions used to represent the distribution function in the velocity phase space.
\\
 \hspace{5mm}\textbullet\ F-(ii) Another misunderstanding about the fluid theory is that the fluid closure appears only due to the fluid reduction of the kinetic Boltzmann equation. This statement is incorrect since the fluid closure appears due to the representation of the distribution function by a finite number of terms. It can be either by the discretization of the distribution function in the velocity phase space, or by the truncation of the number of basis functions used to approximate a continuous distribution function. This means that all existing (gyro)kinetic codes are related to a fluid closure~\footnote{This means that all kinetic and gyrokinetic codes can be exactly reproduced with an equivalent fluid code which contains an extended set of fluid equations similar to the ones shown in Sec.~\ref{ToC:NewFluid}.}.
\\
 \hspace{5mm}\textbullet\ F-(iii) By performing the fluid reduction of the kinetic equation without using a truncation of the representation of the distribution function in the velocity phase space, we obtain the hierarchical fluid equations for an infinite number of fluid moments. By assuming the truncation of the dynamical evolution of a finite number of fluid moments, we obtain an additional unknown fluid moment in the last equation (i.e., $N+1$ unknowns for $N$ equations). This is well known as the problem of the fluid closure and many studies have used a relation of this additional fluid moment as function of the previous fluid moments. We can categorize the different fluid closures found in the literature. The first category is the local collisional fluid closures such as found in Ref.~\citep{Braginskii}. The second is the nonlocal collisional fluid closure such as found in Refs.~\citep{Hammett,Dimits}. The third one is the local collisionless fluid closure such as found in Refs.~\citep{Grad_1949_CPAM_2,Grad_1963_PoF_6} where the distribution function is represented by a truncation of the Hermite polynomials. The later gives the well known $13$ moments fluid model but as explained above, this representation is not efficient for the description of localized deviations of NMDFs. Moreover, another collisionless fluid closure which is not physically correct is found in Ref.~\citep{DeGuillebon:12}, where the authors assume that $M_k=0$ for $k>N$. Their choice is too constraining for non-Maxwellians because even the MDF cannot be described by this method. Indeed, for a MDF only the even moments $\{P_{2k+1}, k\in\mathbb{N}\}$ are canceled and the odd moments $\{P_{2k}, k\in\mathbb{N}\}$ are function of the density, fluid velocity and temperature, and none of the moments $M_k$ are equal to $0$ (the moments $M_k$ and $P_k$ are defined in App.~\ref{ToC.App.dMkdt}).  In the case of the INMDFs given by \refeq{eq.fnM.Generalization}, the collisionless and collisional local fluid closures are obtained by writing the additional moment as function of the hidden variables ($M_k = A u$) instead of the previous moments. In fact, there is no physically motivated reason to have reversible relations between the hidden variables ($u=\{ (a_k,b_k,c_k,d_k,e_k), \textrm{for }k \in \mathbb{N} \}$) and the fluid moments (i.e., such as for the MDF). This non-reversibility is one of the reasons why the collisionless fluid closure of NMDFs has never been developed until now. However, it is important to remark that even if this relation is not reversible, the next generation of fluid code (based on the generalized equations given in Sec.~\ref{ToC:NewFluid}) can evolve the dynamical equations of the hidden variables $u$ by inverting the matrix $B$ where $\partial_{\xi} M_k = B \partial_{\xi} u$ for $\xi$ a space or time coordinate (i.e., $\partial_{\xi} u = B^{-1} \partial_{\xi} M_k$).
\\
 \hspace{5mm}\textbullet\ F-(iv) It has been argued that the fluid theory is valid only for negligible mean-free-path lengths in comparison to all characteristic scale lengths. This means that the fluid theory is only valid when the collision frequency is higher than all characteristic frequencies as explained by the quoted citation above. Hence, one may think that the fluid theory is more valid when the collisionality goes to infinity. However, many experiments are not consistent with this infinite collisionality assumption and the fluid theory can still produce a good representation if we use a consistent set of fluid equations (i.e., at least with our collisionless fluid closure shown in Sec.~\ref{ToC:NewFluid}). Indeed, for an infinite collisionality the distribution function tends to be closer to a MDF but for finite collisionality NMDFs may be observed. The fluid theory is not invalid for a finite collisionality (i.e., departures of Maxwellian distributions when the mean-free-path is not negligible) but the validity of the fluid theory is assured by a large number of particles (in order to have a good statistical kinetic representation), and is particularly related to the representation of the distribution function with continuous basis functions in the velocity phase-space. In other words, the fluid theory is valid when the kinetic theory is. \\

As a conclusion of the limitations of the fluid theory found in the literature, we argue that the description of fluid models including non-Maxwellian kinetic effects even in the collisionless limit has never been obtained self-consistently since 1872 and 1877 when~\citet{Maxwell_1872} and~\citet{Boltzmann_1877} proved that only the MDF is the solution of the Boltzmann equation without external sources. This makes sense because all attempts to represent variations in the velocity phase space of NMDFs were (i) not efficient, reducing to more than a few hundreds or thousands of fluid moments, and (ii) not self-consistent with an intermediate collisionality. The second reason of this failure is the misconception of the fluid reduction from the kinetic Boltzmann description, and more generally of the fluid theory: it has been difficult to expand this fluid theory here for non-equilibria based on the properties of MDFs without falling into the usual limitations, but a summarized point of view of the state-of-the-art of the fluid reduction since the 1870's clarifies our novel approach. \\
The introduced INMDFs represent an extension of the usual fluid theory since (i) it has been shown~\citep{Izacard_2016_Paper1} to fit a numerical NMDF which resolves the TS-ECE discrepancy of the electron temperature measurement in JET~\citep{Luna_2003_RSI,Beausang_2011_RSI}, by using only 6 additional parameters, (ii) the collisionless and collisional fluid closure shown in Sec.~\ref{ToC:NewFluid} prove that the fluid closure is entirely related to the choice of the description of the distribution function in the velocity phase space, and (iii) the INMDF can be consistent with a new solution of the Boltzmann equation for intermediate collisionality regimes in presence of external sources. \\
The INMDF seems to prove that all accepted misunderstandings about the fluid theory mentioned previously can be bypassed.

\subsection{Inconsistency between all fluid models and non-Maxwellians}
We have previously argued that all existing fluid models are related to a MDF. This part explains the reasons of this strong assumption. Starting from the Boltzmann equation and an assumed non-Maxwellian steady-state such as the one measured in experiments or computed from PIC or Fokker-Planck codes, we know that if the NMDF is smooth enough (i.e., continuous and infinitely derivable) then it is a solution of the collisionless limit of the Boltzmann equation (i.e., the \citet{Vlasov_1936} equation $\partial_t f + D\left[\cdots\right] = 0$ where $D\left[\cdots\right]$ is the advection operator described in App.~\ref{ToC.App.dMkdt}). Hence, the collisionless steady-state limit of all fluid models should give us some information about the link between these fluid models and their associated NMDFs.\\
In a general point of view, the fluid models can be wrote in the form
\bgeqa
\label{eq.Lim.dndt.Lit}
\displaystyle
\partial_t n &=& D\left[\cdots\right] + D_n (\cdots) + S_n, \\
\label{eq.Lim.dvdt.Lit}
\displaystyle
\partial_t v &=& D\left[\cdots\right] + D_v (\cdots) + S_v, \\
\label{eq.Lim.dTdt.Lit}
\displaystyle
\partial_t T &=& D\left[\cdots\right] + D_T (\cdots) + S_T,
\edeqa
where $D_{\xi}$ (respectively $S_{\xi}$) are the dissipative (respectively source) terms for $\xi \in \{n,v,T\}$. The fluid hierarchy is such that the equation of the temperature involves the next fluid moment, i.e., the heat flux $q$. Usually, the fluid closure assumes a heat conduction relation
\bgeqa
\displaystyle
q=-\chi \nabla T,
\edeqa
where $\chi$ is the {\color{update}constant} heat conductivity coefficient {\color{update}(other fluid closures are discussed in Sec.~\ref{ToC:limitations.nonlocal} where the local spatial gradient is replaced by a Hilbert or a Fourier transform)}. The ad-hoc dissipative and source terms approximate kinetic effects related to the fluctuations of the NMDF around its Maxwellian steady-state (i.e., it is a perturbation theory). {\color{update}This technique using ad-hoc transport coefficients is not self-consistent because the kinetic effects (approximated by the transport coefficients, see Sec.~\ref{ToC:NewFluid}) are static at the same time that standard fluid quantities (i.e., $n$, $v$ and $T$) are dynamical. Our generalized fluid theory shown in Sec.~\ref{ToC:NewFluid} is self-consistent because the kinetic effects and the associated transport are also dynamically evolving.} However, the steady-state limit of these fluid models given by Eqs.~(\ref{eq.Lim.dndt.Lit})-(\ref{eq.Lim.dTdt.Lit}) becomes
\bgeqa
\displaystyle
\partial_t n &=& D\left[\cdots\right], \\
\displaystyle
\partial_t v &=& D\left[\cdots\right], \\
\displaystyle
\partial_t T &=& D\left[\cdots\right],
\edeqa
with the fluid closure given by
\bgeqa
\displaystyle
q &=& 0.
\edeqa
{\color{update}This fluid closure is the well-known adiabatic (or double-adiabatic, when pressure anisotropy is retained) approximation~\citep{CGL}, and its validity conditions are well known as well, especially at the level of the linear kinetic theory.} All ad-hoc dissipative and source terms are dropped in steady-state because the distribution function is assumed to be Maxwellian. Indeed, it turns out that the steady-state limits of almost all existing fluid models found in the literature are only related to MDFs and are not consistent with the description of non-Maxwellian effects. This means that all non-Maxwellian steady-state distribution functions measured in experiments or computed from PIC or Fokker-Planck codes cannot be described self-consistently and rigorously by any of the existing fluid models. We remark that the most advanced technique helping to better mimic kinetic effects in the state-of-the-art existing fluid codes consist to use radial and poloidal profiles of the transport coefficients $D$ and $\chi$ such as in Refs.\citep{Canik,Groth_2013_NF_53,Meier}. This technique has been developed to match midplane profiles of density and temperature obtained from experimental measurements in expectation to better describe the dynamics of the SOL and the divertor plasma. It has been found that this profile fitting technique reduces the radiation shortfall~\citep{Groth_2013_NF_53} and can help to simulate the transport enhancement between the two X-points of a snowflake magnetic configuration~\citep{Rognlien_2014_APS,Izacard_2016_APS}. However, even if this technique allows to describe different transport levels at different positions, these transport coefficients are assumed ad-hoc and there is no direct relation with the distribution function. As described below, another extension to these fluid models is the use of nonlocal transport coefficients.

\subsection{Non-self-consistent solution: nonlocal heat transport and others}
\label{ToC:limitations.nonlocal}
One of the most advanced attempt to approximate some kinetic effects in the fluid equations is to use the nonlocal heat transport theory. Instead of using the usual fluid closure $q=-\chi \nabla T$ where $\chi$ is an ad-hoc coefficient constant in time, fixed in order to match some experimental data, in some cases only the use of a nonlocal definition of the dissipative coefficients can enhance the validation of simulations against experiments~\citep{Dimits} and can be consistent with some kinetic effects~{\color{update}\citep{Hammett,Chang_1992_PoFB,Snyder_1997_PoP}}. It turns out that this choice uses the linearization of the Fokker-Planck equation (similar to Ref.~\citep{Ji_2013_PoP_13}), a {\color{update}Hilbert or a} Fourier transform, and is always associated with fluctuations of the distribution function. A nonlocal heat flux is defined by $q = \int F[n,v,T] d^3x$ where $F$ is an ad-hoc functional of the fluid quantities (usually motivated by the description of the phenomenon under investigation). This nonlocal transport theory can be robust because the nonlocal definition efficiently adds a time evolution of the dissipative transport coefficients as function of the fluid quantities, but there is no robust and simple link between a specific shape of the NMDF and the ad-hoc functional $F$. {\color{update}Similarly, more advanced fluid closures at higher order ranks have been proposed~\citep{Snyder_1997_PoP,Scott_2010_PoP,Sulem_Passot_2014_JPP} but they always involve perturbations of the distribution function. The fluid closure of Ref.~\citep{Grasso_2015_JPP} is obtained by adding strong constraints from Ref.~\citep{Scott_2010_PoP} due to the interest of the authors in finding the Hamiltonian structure~\footnote{{\color{update}We note that the Hamiltonian structure of the collisionless limit of our generalized fluid theory detailed in Sec.~\ref{ToC:NewFluid}.1 is intrinsically conserved from Maxwell-Vlasov by the change of variable from the distribution function to the hidden variables.}}.} Even though the nonlocal transport is a way to describe modifications of the anomalous transport due to turbulence and possibly non-Maxwellian kinetic effects, this nonlocal theory is still a perturbation theory and is not consistent with any steady-state NMDF which should naturally include a local closure as shown in Sec.~\ref{ToC:NewFluid}.

\section{First fluid models consistent with non-Maxwellian steady-states}
\label{ToC:NewFluid}
A robust new set of fluid equations can be constructed. This new kind of fluid equations associated with NMDFs has to: (i) be linked with collisionless fluid closures, (ii) describe departures to the usual fluid models related to a MDF in the steady-state limit, (iii) use local collisional fluid closures which include space and time variations. The development of these fluid models is reported here.
\subsection{Local collisionless fluid closures}
The fluid equations derived from the moments of the Vlasov equation (i.e., the collisionless limit of the Boltzmann equation), are wrote in a much compact form using the moments $M_k$ rather than the moments $P_k$ (see definitions in App.~\ref{ToC.App.dMkdt}). The fluid equations (previously noted $\partial_t M_k = D\left[\cdots\right]$) read
\bgeqa
\displaystyle
\partial_t M_0 &=& - \nabla \cdot M_1, \\
\displaystyle
\partial_t M_k &=& - \nabla \cdot M_{k+1} - \frac{e}{m} \left( M_{k-1} {\bf E} + M_k \times {\bf B} \right),
\edeqa
for $k \in \mathbb{N}^{\star}$. By assuming that the distribution function is described by the INMDF given by \refeq{eq.fI}, all fluid moments $M_k$ are function of the $6$ hidden variables $(n,v,T,\Gamma,c,W)$ (see App.~\ref{ToC.App.Mk_INMDF}), specially the moment $M_{6}$ which appears in the last dynamical moment equation $\partial_t M_5$. The collisionless fluid closure associated to the INMDF~\citep{Izacard_2016_FischSymposium,Izacard_2016_Paper1} given by
\bgeqa
\label{eq.fI}
f_I &=& \Delta \exp\left(-\frac{1}{2T} {\rm v}^2 + \frac{v}{T} {\rm v} \right) + \Delta_I ({\rm v}-c) \exp\left(-\frac{1}{2W} {\rm v}^2 + \frac{c}{W} {\rm v} \right),
\edeqa
where $\Delta=n (2\pi T)^{-1/2} \exp(-v^2/(2T))$ and $\Delta_I=\Gamma (2\pi W^3)^{-1/2} \exp(-c^2/(2W))$, is
\bgeqa
\label{eq.M6}
\displaystyle
M_6 &=& n \left( 15T^3+45T^2v^2+15Tv^4+v^6 \right) 
+ \Gamma c \left( 15W^2+10Wc^2+c^4 \right).
\edeqa
{\color{update}The physical interpretation of the hidden variables $(n,v,T,\Gamma,c,W)$ is given by~\citet{Izacard_2016_Paper1}.} From the point of view of numerical codes, we have to remark that an algorithm of each time step can be constructed as the following:\\
 \hspace{5mm}\textbullet\ A-(1) Each time step of the generalized fluid code needs inputs. For each time step except the initial condition, the fluid hidden variables $u=(n,v,T,\Gamma,c,W)$ associated with the INMDF are given. Only the initial time step may be different: we can use a numerical distribution function $f_{init}$ or the numerical values of its $6$ first fluid moments $M=(M_0,M_1,M_2,M_3,M_4,M_5)$. In both cases, a fitting algorithm is required in order to match $f_{init}$ or $M$ with the analytic computation assuming the INMDF and the fitted initial conditions of the hidden variables $u$.
\\
 \hspace{5mm}\textbullet\ A-(2) The second step of this algorithm is to numerically compute the RHS of the dynamical equation of the fluid moments $M$ (see App.~\ref{ToC.App.dMkdt}). For that, we need the fluid moments $M$ as function of the hidden variables $u$ (see App~\ref{ToC.App.Mk_INMDF}) $M = A u$ where $A$ is a linear matrix function of the hidden variables $u$, as well as the spatial derivatives of the fluid moments $\partial_{\xi} M_k = B \partial_{\xi} u$ where $\xi \in \{x,y,z\}$ and $B$  is a simple matrix function of the hidden variables $u$. In other words, we compute the RHS of the dynamical equations of the fluid moments $\partial_t M$ as function of the hidden variables $u$ and its spatial derivatives.
\\
 \hspace{5mm}\textbullet\ A-(3) The third step of this algorithm is to invert the matrix $B$ (function of $u$) which also links the time derivatives $\partial_t M = B \partial_t u$. Then,  $\partial_t u = B^{-1} \partial_t M_k$ where $B^{-1}$ is a simple matrix which is a function of the hidden variables $u$.
\\
 \hspace{5mm}\textbullet\ A-(4) The last step is the computation time evolution of the hidden variables $u$ using one of the available numerical schemes. Then, the $6$ fluid hidden variables $u$ are updated and the loop of the algorithm is closed. 
\\
An important remark is the fact that the closure on the moment $M_6$ is part of the step A-(2) in this algorithm because the fluid moment $M_6$ is known as function of the hidden variables $u$.\\
At this point, we have introduced here $6$ fluid equations associated with the INMDF whereas the current fluid theory is {\color{update}mainly} based on the common $3$ fluid equations associated with MDFs {\color{update}(see Sec.~\ref{ToC:limitations.nonlocal} for discussion on exceptions)}. However, it is possible to reduce the degrees of freedom of the INMDFs by assuming $2$ constraints relevant to physical motivations of the system under consideration. As an example, it is possible to use $2$ constraints if one wants to describe INMDFs where the central flow $c(n,v,T,\Gamma)$ and the width of the heat spread $W(n,v,T,\Gamma)$ are given as function of the $4$ dynamically independent hidden variables $u=(n,v,T,\Gamma)$. As an example from the construction of the coefficients $r$ and $s$ in Ref.~\citep{Izacard_2016_Paper1} where $c=v+r\sqrt{T}$ and $W=s^2T$, we can constraint a specific value for the coefficient $r$ and $s$ in order to remove the independence of $c$ and $W$ from the $4$ other fluid hidden variables. In this case, the collisionless fluid closure that appears in the dynamical equation $\partial_t M_3$ is given by
\bgeqa
\label{eq.M4}
\displaystyle
M_4 &=& n \left( 3T^2+6Tv^2+v^4 \right) + 4\Gamma c \left( 3W+c^2 \right),
\edeqa
where the constraints on $c(n,v,T,\Gamma)$ and $W(n,v,T,\Gamma)$ are required here. \\
These two $4$-moments and $6$-moments fluid models are the first ones efficiently related to a NMDF that can describe a tail or a disymmetry. Both models are directly reduced from the INMDF. A large number of reduced models can be obtained from this set of fluid equations and the perturbative theory may help in adding more terms related to turbulence, in the same way it has been developed in the previous decades for fluid models close to MDFs. Indeed, we can construct fluid models for turbulence (i.e., fluctuations of the distribution function) close to an INMDF. As shown for example for the characteristic curve of the Langmuir probe by \citet{Izacard_2016_FischSymposium,Izacard_2016_Paper1} where the INMDF was able to replace an ad-hoc diffusion term, a population of super-thermal particles can naturally enhance the transport. It is then trivial to model fluctuations by a sum of INMDF deformations of the distribution function. We might be able to recover the anomalous transport due to the presence of turbulence by only considering NMDFs. {\color{update}Moreover, the dissipative terms present in the turbulence models can be recovered by specific constraints of the hidden variables. For example, it turns out that if we impose the kinetic flux $\Gamma = - D \nabla n$, the divergence of the first moment $M_1$ which appears in the continuity equation $\partial_t n$ becomes $- \nabla \cdot M_1 = - \nabla \cdot (nv) + \nabla \cdot ( D \nabla n)$. Then, if the diffusion coefficient is homogeneous, we recover the usual particles diffusion $- \nabla \cdot \Gamma = D \nabla^2 n$. Another interesting observation is found when we assume the constraints $v=c=0$ and constant $T$ and $W$. We then obtain from the dynamical moments equations $\partial_t M_k$ for $k \in \left[0,1\right]$ (see the moments in Appendix~\ref{ToC.App.Mk_INMDF}) of an INMDF the two following equations
\bgeqa
\label{eq.INMDF.Diffusion.n}
\displaystyle
\partial_t n &=& - \nabla \cdot \Gamma,\\
\label{eq.INMDF.Diffusion.Gamma}
\displaystyle
\partial_t \Gamma &=& - T \nabla n.
\edeqa
Then, with the linearlization of the term $\partial_t \Gamma$, we find the relation $D=T/\gamma$ where $\gamma$ is the linear growth rate of the kinetic heat flux $\Gamma$. In other words, the diffusion coefficient $D$ has its origin in a static linear growth rate of the kinetic heat flux.
\begin{figure}
\centerline{
(a)\label{fig.Diffusion.test}
\includegraphics[height=45mm]{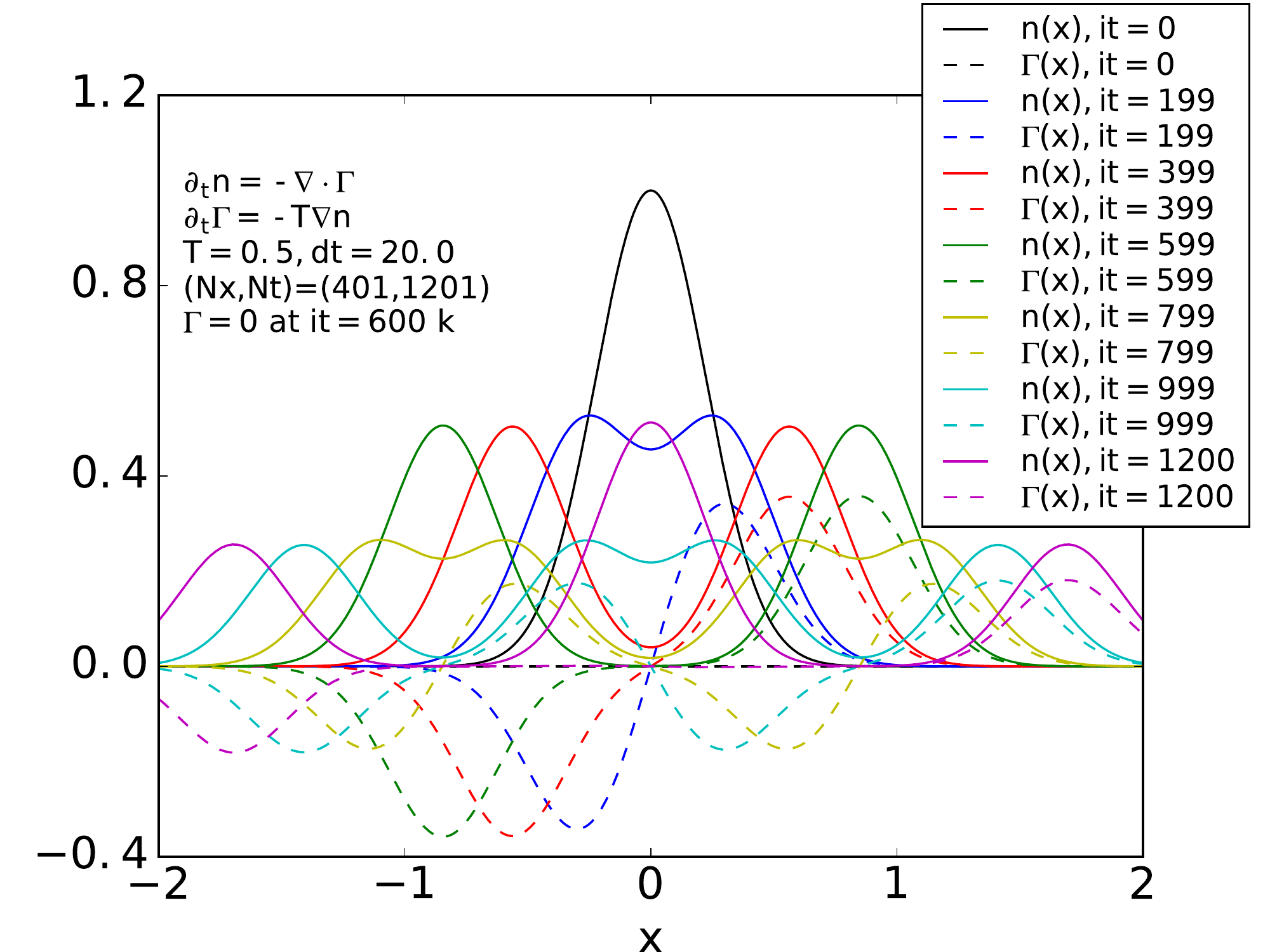}
(b)\label{fig.Diffusion.adaptative}
\includegraphics[height=45mm]{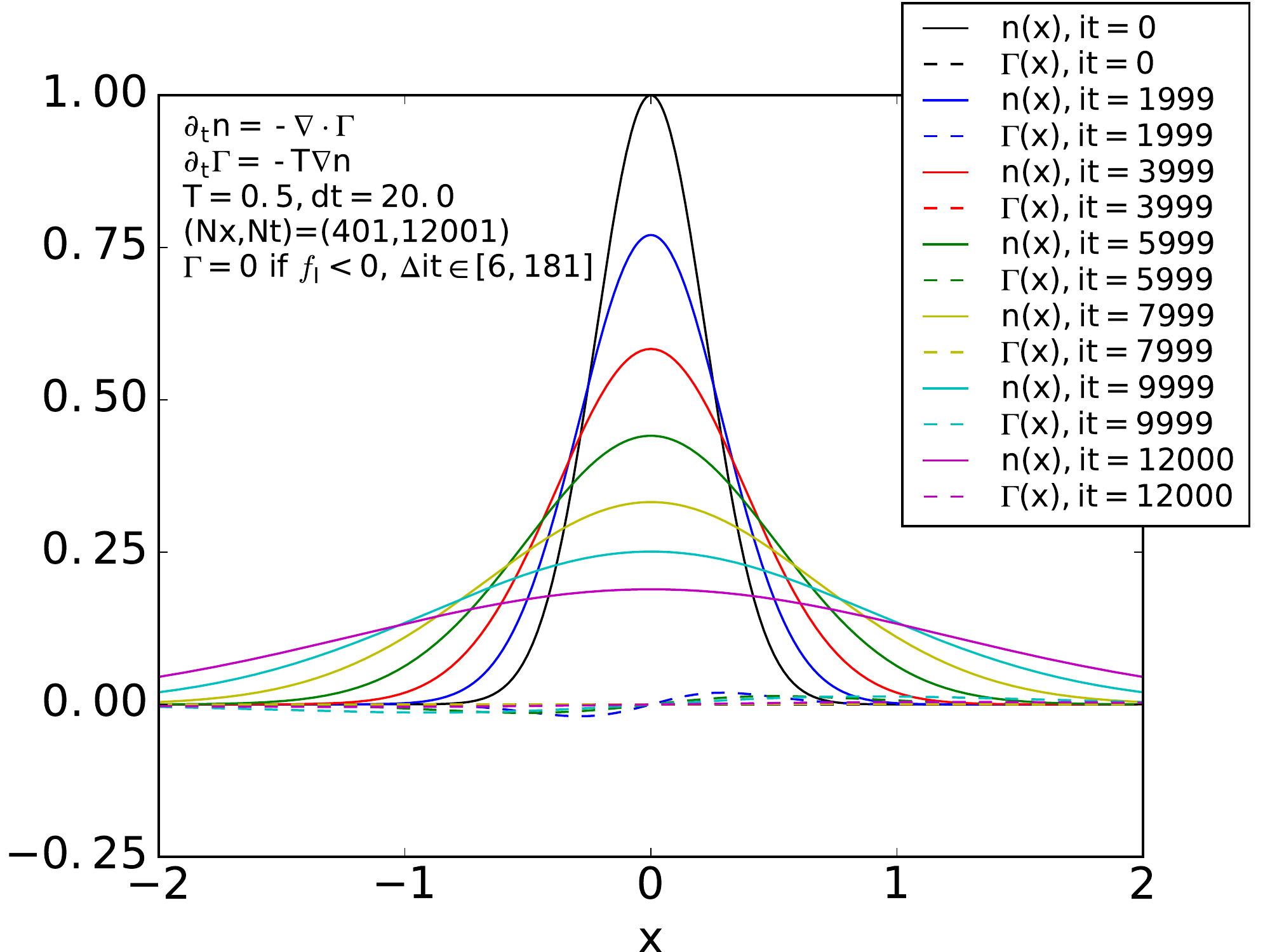}
}
\caption{{\color{update}Validation studies of Eqs.~(\ref{eq.INMDF.Diffusion.n})-(\ref{eq.INMDF.Diffusion.Gamma}) proving the kinetic origin of the particle diffusion via the competition between the growth of the INMDF given by \refeq{eq.fI} and the thermalization relaxation simulated by forcing $\Gamma=0$ at specific times.}}
\label{fig.Diffusion}
\end{figure}
As a validation test, Fig.~\ref{fig.Diffusion} shows two numerical simulations of the reduced INMDF fluid model given by Eqs.~(\ref{eq.INMDF.Diffusion.n})-(\ref{eq.INMDF.Diffusion.Gamma}). The initial value of the density $n(x,it=0)$ is a Gaussian function (solid black curve) and that of the kinetic heat flux $\Gamma(x,it=0)$ is $0$ (dashed black curve). Moreover, the non-periodic spatial gradients are evaluated with the \texttt{numpy.gradient} function in Python and the time integration with a first order difference scheme. If we let the system $(n,\Gamma)$ evolve in time, we observe in Fig.~\ref{fig.Diffusion}.(a) the splitting of the Gaussian of the density in two propagating Gaussians in opposite directions (blue, red and green curves) and the propagating velocity is a directly function of the constant temperature $T=0.5$ (we also fixed $W=T/2$ for the INMDF). However, after $600$ iterations in time (at $t=it \times dt$, where $it=600$ and the time step $dt=20$ here) we chose to manually reset $\Gamma(x,it=600)$ to $0$ because in this reduced INMDF fluid model there is no collision that assures a positive distribution function. This manual reset simulates the thermalization of the INMDF toward the MDF (when $\Gamma=0$). We note the effect of resetting $\Gamma(x,it=600)=0$ (green curves) in Fig.~\ref{fig.Diffusion}.(a): each Gaussian splits in 2 new Gaussians propagating again in opposite directions (yellow, cyan and magenta curves). However, it turns out that due to the high value of the time step and spatial gradients, as well as our choice of the temperature $T$ and width of heat spread $W$, the distribution function becomes negative after a few time steps and its minimum value saturates when the dynamics becomes purely advective. We remark that other choices of $T$ and $W$ would have resulted in a saturated advection with a positive INMDF but the dynamics would have been slower. Fig.~\ref{fig.Diffusion}.(b) is the same simulation but the kinetic heat flux $\Gamma$ is reset to $0$ at every time step before getting a negative distribution function (see criteria in Ref.~\citep{Izacard_2016_Paper1}). The finite collisionality is simulated by the duration $\Delta it$ of the thermalization which varies from $6$ (at $it=0$) to $181$ (at $it=12000$) time steps due to the reduction of the spatial density gradients by the diffusion~\footnote{{\color{update}Other simulations (not discussed here) have been performed with spatial profiles of the density with nonphysical large gradients. In this case, the time step $dt$ must be reduced in this case in order to avoid negative INMDFs.}}. As a summary of these reduced INMDF fluid numerical simulations, we prove here the origin of the diffusion via kinetic effects (i.e., via the existence of the INMDF due to spatial gradients) on shorter time scales than those of the thermalization. We obtained similar numerical results for the heat conduction with $\partial_t (nT)=-3 \nabla \cdot (\Gamma W)$ and $\partial_t ( \Gamma W )=-\nabla (n T^2)$ where we can choose constant and homogeneous $n$ and $W$. These results are unique thanks to the use of the INMDF since by using the momentum $nv$ of a MDF instead of $\Gamma$ there is no physical motivation to reset the momentum to $0$ on the time scales of the thermalization. These observations prove that (i) the kinetic heat flux exists at least in part due to the existence of spatial density gradients, (ii) the usual static particle diffusion and heat conduction are not consistent with dynamical density and temperature gradients (and with our dynamical growth rate of the kinetic heat flux), and (iii) the INMDF has a physical existence via our understanding of the origin of particle and heat diffusion~\footnote{{\color{update}In addition to the physical existence of the INMDF proved by the fitting of the experimental observation of a NMDF bulk in JET~\citep{Izacard_2016_Paper1}.}}. 
These significant observations based on very simple INMDF fluid models are the keys that allow us to be optimistic for a potential understanding of the turbulence transition. These results are more discussed in the perspectives and will be the purpose of future publications.}\\
Until now we have explained only the collisionless fluid closure. The next step is to introduce the collisional fluid closure that is responsible of the relaxation toward the MDF equilibrium.

\subsection{Local collisional fluid closures}
There are two types of collisional fluid closures: the self-collisions of the particle species with themselves and the collisions with heated particles coming from a source of energy (e.g., NBI, radiofrequency waves, alpha particles, other species). The former tends to reduce the deviation from the MDF which is the thermodynamic equilibrium without sources, but the presence of the later constantly generates deviations from the MDF and a steady-state NMDF is possible at finite collisionality. Here we argue that this steady-state can be described by the kinetic theory as well as by the generalized fluid theory introduced by our work. \\
For the first type of collisional fluid closure, the nonlinear Fokker-Planck collisional operator is required for the description of low angle scattering by two-body collisions. By assuming that the steady-state distribution function is an INMDF $f_I$ instead of a MDF, we compute the moments of the self-collision Fokker-Planck operator
\bgeqa
\displaystyle
C_k = \int_{-\infty}^{\infty} C[f_I,f_I] {\rm v}^k d{\rm v},
\edeqa
and the fluid equations read
\bgeqa
\displaystyle
\partial_t M_0 &=& - \nabla \cdot M_1 + C_0, \\
\displaystyle
\partial_t M_k &=& - \nabla \cdot M_{k+1} - \frac{e}{m} \left( M_{k-1} {\bf E} + M_k \times {\bf B} \right) + C_k,\qquad
\edeqa
where $C_k$ are function of hidden variables of the INMDF. We found~\citep{Izacard_2016_FischSymposium} that the terms $C_k$ introduce a new function $I_k(a,b)$ called the plasma collision function that is defined by
\bgeqa
\displaystyle
I_k(a,b) = \int_{-\infty}^{\infty} x^k \sqrt{1+x^2} \exp(-ax^2+bx) dx.
\edeqa
In fact, the nonlinear Fokker-Planck collision operator can be written in a form involving only one Rosenbluth potential~\citep{Rosenbluth} as shown by~\citet{Gaffey_1976_JPP_16}. The correction of the Rosenbluth potential from $F(x) = 1/(nv) \int_{-\infty}^{\infty} \vert {\bf v}_i-{\bf v} \vert f({\bf v}) d^3{\bf v}$ where $x={\rm v}_i/v$ for a MDF $f=f_0$ (i.e., the correction from $F_0(x)=(x+1/(2x)) {\rm Erf}(x)+\exp(-x)/\sqrt{\pi}$) introduces the plasma collision function $I_k(a,b)$
\bgeqa
\displaystyle
F_I(x) = F_0(x) + \Delta_I {\rm v}_{i\perp}^2 \bigg[ {\rm v}_{i\perp} I_1\left(A,B\right) + ({\rm v}_{iz}-c) I_0\left(A,B\right) \bigg],
\quad\edeqa
where $A={\rm v}_{i\perp}^2 / (2W)$ and $B={\rm v}_{i\perp}(c-{\rm v}_{iz}) / (2W)$. Moreover, even an anisotropic MDF (e.g., with different perpendicular $T_{\perp}$ and parallel $T_{\parallel}$ temperatures) introduces this plasma collision function. In order to obtain the correction $F_I(x)$, an anisotropic 3D version~\citep{Izacard_2016_FischSymposium} of the INMDF is required. We remark that the tensor version of the fluid equations are given in App.~\ref{ToC.App.dMkdt}. More details of the development of the nonlinear Fokker-Planck collision operator from a 3D version~\citep{Izacard_2016_FischSymposium} of the INMDF will appear in a future publication since the focus of this work is to understand the limitations of the current fluid theory and to explain the key points allowing the self-consistent generalization of the fluid theory including kinetic effects. \\

For the second type of collisional fluid closure, we need to add as a source or sink of energy, the collision with other species. The distribution function of these other species can be assumed or evolved by the Fokker-Planck equation. In the later case, if there are no external sources in both Fokker-Planck equations, both steady-states can only be Maxwellians and are not relevant to experiments. However, for example, the addition of a NBI in one or the other Fokker-Planck equations is the key to obtain NMDFs steady-states for both species. The source term, by assuming an INMDF $f_I$ as a first species and a source $f_s$ as a second species, is defined by
\bgeqa
\label{eq.Closure.Sk}
\displaystyle
S_k = \int_{-\infty}^{\infty} C[f_I,f_{s}] {\rm v}^k d{\rm v},
\edeqa
and the fluid equations become
\bgeqa
\displaystyle
\partial_t M_0 &=& - \nabla \cdot M_1 + C_0 + S_0, \\
\displaystyle
\partial_t M_k &=& - \nabla \cdot M_{k+1} - \frac{e}{m} \left( M_{k-1} {\bf E} + M_k \times {\bf B} \right) 
+ C_k + S_k.
\edeqa
Following the techniques published by~\citet{Gaffey_1976_JPP_16}, it could be possible to derive the steady-state distribution function $f_s$ of a NBI as a response to the INMDF, and then to compute the Fokker-Planck collision operator and the source term given by \refeq{eq.Closure.Sk}. From this computation, we should be able to check if the INMDF is the first analytic self-consistent solution of the Boltzmann equation in presence of sources such as NBI. \\

In summary, we have demonstrated the method to construct generalized fluid models including kinetic effects. We have made the distinction between the collisionless and collisional fluid closures which are both local and defined by the velocity phase space representation of the distribution function. For the next generation of fluid codes, we have shown an algorithm which evolves the dynamics of the fluid hidden variables, defining both the distribution function and the fluid moments. The unification between kinetic and fluid theories is obtained. 
The highlight on some results and a list of perspectives are detailed in Sec.~\ref{ToC:CCL}.

\section{Discussions and future work}
\label{ToC:CCL}
This report introduces a novel description of NMDFs and gives a global point of view of the plasma physics theory and especially the fluid reduction. It is specially important to highlight some results and to list the most important perspectives.\\ 
 \hspace{5mm}\textbullet\ H-(i) The INMDF and some generalizations are introduced here and in Refs.~\citep{Izacard_2016_FischSymposium,Izacard_2016_Paper1} as function of hidden variables for the analytic description of NMDFs. The fluid moments are simply obtained as function of the hidden variables. Our use of the hidden variables is very efficient for the statistical description of non-isolated systems where sources and sinks of energy are possible without the necessity to include the exact description of all processes interacting in or with the system.
\begin{figure}
\centerline{
\includegraphics[height=65mm]{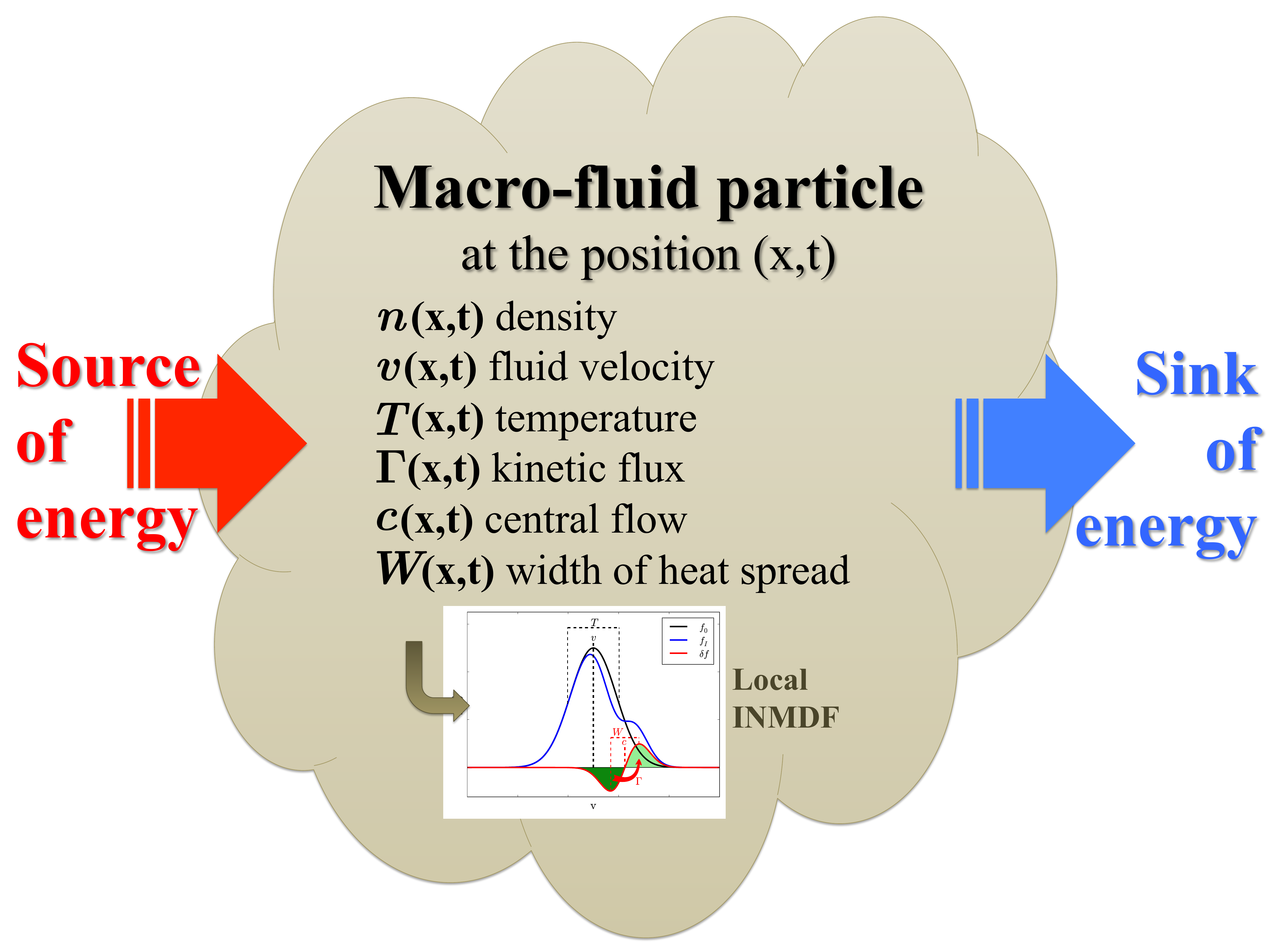}}
\caption{Scheme of a macro-fluid particle corresponding to a statistical distribution represented by the simplest INMDF for a non-isolated system. The source and sink of energy are local for this macro-fluid particle. The macro-fluid hidden variables: the density $n(x,t)$, the fluid velocity $v(x,t)$, the temperature $T(x,t)$, the kinetic flux $\Gamma(x,t)$, the central flow $c(x,t)$ and the width of the heat spread $W(x,t)$ characterize the statistical representation of particles in presence of sources and sinks.}
\label{fig.CCL.MacroFluidParticle}
\end{figure}
Fig.~\ref{fig.CCL.MacroFluidParticle} represents a scheme of a macro-fluid particle in presence of sources and sinks of energy such that the distribution function is represented by an INMDF. {\color{update}We choose the name of ``macro-fluid particle" instead of the usual ``fluid particle" which appears in the literature in order to highlight the fact that the statistical internal properties (i.e., the distribution function) is not thermalized due to the presence of sources and sinks of energy at finite collisionality.} The Eulerian fluid quantities are the hidden variables, physically interpreted in Refs.~\citep{Izacard_2016_FischSymposium,Izacard_2016_Paper1}. {\color{update}We remark that our generalized fluid theory based on nonorthogonal basis sets is efficient because we used a similar approach than the collective coordinates of the Rayleigh-Ritz variational method which reduces the number of degree of freedom of a system by using collective coordinates.}
\\
 \hspace{5mm}\textbullet\ H-(ii) Additional corrections due to super-thermal particles are detailed here such as a modified nonlinear Fokker-Planck collision operator. It is argued that INMDFs are the statistical description of non-isolated plasmas and is the first possible analytic solution of the Boltzmann equation which deviates from the isolated MDF developed in the 1870's. This result demystifies the Maxwell's demon, not by continuously including more external processes in order to try to describe an isolated system.
\\
 \hspace{5mm}\textbullet\ H-(iii) After some explanations of the failure to efficiently describe non-Maxwellian effects in the fluid models found in the literature, the first self-consistent generalized fluid models compatible with super-thermal INMDFs are given as well as the collisionless and collisional fluid closures. The formulation of the nonlinear Fokker-Planck collision operator as function of the hidden variables helps to include self-consistent transport due to kinetic effects in the new fluid equations. {\color{update}Moreover, we show here the collisionless analytic constraints linking kinetic NMDFs and the ad-hoc transport coefficients (particle diffusion, viscosity and heat conductivity) of the standard fluid theory. However, without constraints, our new generalized fluid theory obtained from analytic NMDFs does not required ad-hoc transport coefficient since the dynamical transport is already self-consistently included by the additional fluid quantities (e.g., kinetic heat flux, central flow, and width of the heat spread for the INMDF).}
\\
These results {\color{update}would lead to groundbreaking results for} plasma physics theory by the simplicity of the fluid equations obtained without using most of the common assumptions or complex methods for the description of NMDFs. In other words, we built here the missing link between the kinetic and fluid theory since we show the trivial inclusion of kinetic effects in a small number of fluid equations. As a conclusion, we are predicting more accurate results of the next generation of fluid codes based on the equations and ideas developed here in comparison to all kinetic and gyrokinetic codes which are very CPU-consuming. The reason is that all kinetic and gyrokinetic codes use discretized distribution functions, or one of the orthogonal basis sets for the description of NMDFs. These representations has to be compared with the compact, continuous and efficient representations of the mesh-free distribution functions based on non-orthogonal basis sets allowing the efficient and exact fluid reduction. Finally, our description allows us to take into account statistical representations of non-isolated systems relevant to a finite collisionality.\\

These results give us many possibilities to improve theoretical and numerical studies for the plasma physics and fluid theories and, at least, the prediction of fusion energy production by magnetized confinement devices. Here is a list of a few perspectives:\\
 \hspace{5mm}\textbullet\ P-(i) The expansion of the gyrokinetic equations for non-Maxwellian steady-state backgrounds could dramatically enhance the accuracy of transport and turbulence predictions of burning plasmas. With these very large gyrokinetic simulations assuming a MDF and a neoclassical population, it has been possible to better simulate transport and turbulence, but the results on transport can be within factors of $3$ with respect to experimental measurements of the transport. For example, the gyrokinetic turbulence code GYRO is used to numerically compute transport coefficients in short time scales turbulence. The use of these transport coefficients in the fluid transport code TGYRO accurately predicted profiles of the plasma in some regimes such as low confinement regimes in the DIII-D tokamak. For other regime, a recurrent discrepancy is observed. However, it is not self-consistent to describe steady-state deviations from the MDF background (i.e., a tail) with the perturbation theory. Indeed, by definition, fluctuations have a null time average and cannot contribute to the development of deviations of Maxwellian steady-states. This statement is consistent with the additional inclusion of the neoclassical regime (i.e., steady-state trapped particles) in actual gyrokinetic codes. Then, since super-thermal particles can exist for many reasons (e.g., the alpha particles produced by fusion, the heating by radiofrequency waves or neutral beams or the presence of X-points maintained by the external magnetic coils and generated by MHD dynamics), it is required to include a non-Maxwellian background distribution function in these gyrokinetic codes. This can be done numerically but the introduction of the INMDF here could allow in the future to extend the gyrokinetic equations for the description of super-thermal population of particles in a very efficient way. Indeed, we can avoid using one of the orthogonal basis sets which are irrelevant for continuous and local deviations from MDFs. The basis functions introduced above could considerably reduce the computer consumption by using much less hidden variables in these already very CPU-consuming gyrokinetic codes.
\\
 \hspace{5mm}\textbullet\ P-(ii) The development of the next generation of fluid codes could allow a better prediction of burning plasmas confinement in different collisionality regimes. This is especially relevant for spherical-tokamaks such as NSTX-U at Princeton where a lower collisionality than in tokamaks is observed. In fact, the past 20 years have been consequently dedicated to the development of gyrokinetics codes running on supercomputers for the simulation of the plasma on short time scales. The accuracy of those very CPU-consuming codes cannot be verified against other kinetic codes since the computational time needed is not manageable and the keys to the non-Maxwellian fluid models had never been found until the current work. The fluid reduction from the representations of NMDFs based on orthogonal basis sets are clearly associated with too many fluid equations. Our novel approach that considers local deviations of the distribution function in the velocity phase-space using the smallest number of terms, allows the efficient description of NMDFs by fluid equations. In conclusion, significant advances are perceptible in the sense that: First, these next generation fluid codes would have a better accuracy than some existing gyrokinetic and kinetic codes which use discretized distribution functions in the velocity phase space or non-adapted orthogonal basis functions. Second, the next generation fluid codes would use much less numerical resources and would allow for the full integrated simulation of a burning plasma in a fusion reactor on time scales relevant to the efficiency and cost of power plants.
\\
 \hspace{5mm}\textbullet\ P-(iii) Another significant perspective is possible with the universality of non-equilibrium processes. This report focuses on the statistical description of a plasma when super-thermal particles need to be taken into account. This population of particles is the equivalent of non-equilibrium statistics which appear in many areas that could be enhanced using the development shown here such as: In astrophysics where NMDFs are often observed (i.e., solar wind, magnetosphere, magnetic reconnection). In hydrodynamics (i.e., weather prediction, turbulence around planes, liquid dynamics) where the observed turbulence is approximated with ad-hoc terms in the fluid equations, but as shown here, the simple existence of super-thermal particles are directly linked with these ad-hoc dissipative terms. Moreover, the unsolved Navier-Stokes problem could possibly be solved by writing the dissipative ad-hoc terms as function of hidden variables of the distribution function and obtaining analytic solutions. In molecular and chemical dynamics where the reaction processes are modified by the presence of non-equilibrium phenomena. Finally and more generally, in statistics where the presence of a biasing on the statistical phenomenon under investigation can be described with non-equilibrium distribution constructed here with a very small number of hidden variables. We remark that the goal of these perspectives is not to develop a complete theory explaining the reasons of the non-equilibrium processes, but as a statistical theory, we may perform computations, validations and predictions from these non-equilibria. All of these areas could be easily enhanced by the introduction of non-equilibrium processes based on our analytic description.
\\
 \hspace{5mm}\textbullet\ P-(iv) Can we explain the origin of the diffusion and the turbulence by spatial propagation of local kinetic effects? This perspective is directly motivated by this work because it is possible to describe NM fluctuations of the distribution function by small localized displacement of population of particles in the velocity phase space using hidden variables. Additionally, the turbulence is described by the use of the sources and dissipative terms in the fluid equations. Since we explained here that sources and dissipative terms can be directly computed as function of the hidden variables (instead of ad-hoc coefficients), the diffusion and turbulence can be viewed as a spatial propagation of the NM fluctuations of the distribution function. {\color{update}Finally, the self-consistency is reached because} these fluctuations can appear due to the presence of spatial inhomogeneities and external sources.
\\
 \hspace{5mm}\textbullet\ P-(v) We remark the fact that we introduced here a new method to describe the distribution function with a free-velocity-space-mesh and physical interpretation of non-Maxwellians with as few hidden variables as possible, thanks to the use of non-orthogonal basis sets. The neoclassical trapped particles, the fluctuations (i.e., the turbulence), and other deviations of MDFs can use different sets of basis functions than the one shown in Fig.~\ref{fig.Mk.BasisINMDF}. The method seems universal since it can be applied in many other areas than laboratory plasma physics. Moreover, we also choose to use the Fokker-Planck collision operator which is valid only for small angle scattering collisions, but nothing seems to restrain us to choose other collision operators. \\

In summary, we clarify here the limitation of the current fluid theory to include kinetic effects. The introduced generalized fluid theory which efficiently includes some kinetic effects is valid when the kinetic theory is valid. {\color{update}This work strongly suggests that the key point to efficiently unify fluid and kinetic} theories is the use of non-orthogonal basis sets in order to considerably reduce the number of fluid moments necessary to describe the kinetic effects under investigation. The method developed here might also serve gyrokinetic simulations by reducing their numerical cost. Finally, the community could develop these perspectives the same way it did develop plasma physics based on the thermodynamic Maxwellian equilibrium in the last century. This could possibly allow new understandings and predictions for the dynamics of burning plasmas in fusion reactors.

\section*{Acknowledgements}
\addcontentsline{toc}{section}{Acknowledgements}
{\color{update}This work was supported by the 25\% LLNL institutional funding for postdoctoral researchers.} This work was performed under the auspices of the U.S. Department of Energy by Lawrence Livermore National Laboratory under Contract DE-AC52-07NA27344. {\color{update}The author is grateful for the fruitful reviewer's comments.}

\appendix
\addcontentsline{toc}{section}{Appendix}

\section{Tensorial fluid equations}
\label{ToC.App.dMkdt}
From the fluid reduction of the Vlasov equation, the tensorial version of the fluid equations are given by
\bgeqa
\displaystyle
\partial_t M_0 &=& - \nabla_{\alpha} M_{1}^{(\alpha)}, \\
\displaystyle
\partial_t M_k^{(i_1,\cdots,i_k)} &=& - \nabla_{\alpha} M_{k+1}^{(\alpha,i_1,\cdots,i_k)} \nonumber\\
\displaystyle
&&
+ \frac{e}{m} \Big( E^{(i_1)} M_{k-1}^{(i_2,\cdots,i_k)} 
\Big. 
+ \epsilon_{i_1\alpha\beta} M_k^{(\alpha,i_2,\cdots,i_k)} B^{(\beta)}  + \underbrace{\circlearrowright}_{(i_k)}
 \Big), 
\edeqa
where the summation over repeated indices is used, $\epsilon_{i_1\alpha\beta}$ is the Levi symbol, the symbol $\underbrace{\circlearrowright}_{(i_k)}$ stands for the addition of all terms obtained by the cyclic permutation of the indices $(i_1,\cdots,i_k)$, and the moments are $M_k^{(i_1,\cdots,i_k)} = \int_{-\infty}^{\infty} f({\bf v}) \Pi_{j=1}^k {\rm v}^{(i_j)} d^3{\bf v}$. The 3D version of the INMDF~\citep{Izacard_2016_FischSymposium} given by \refeq{eq.fI} will be given in a later publication. The readers can also convert these equations with the moments $M_k$ in order to use the well known moments $P_k^{(i_1,\cdots,i_k)} = 1/M_0 \int_{-\infty}^{\infty} f({\bf v}) \Pi_{j=1}^k \left({\rm v}^{(i_j)}-M_1^{(i_j)}/M_0\right) d^3{\bf v}$. {\color{update}We remark that this work is mainly based on 1D NMDFs but it is straightforward to use this tensorial version of the fluid equations by a multiplication of three 1D NMDFs and Appendix~\ref{ToC.App.Mk_INMDF}. We do not report the final equations because there are many choices of 3D NMDFs based on the nonorthogonal basis given by \refeq{eq.fnM.Generalization} or others.}

\section{Fluid moments of INMDFs}
\label{ToC.App.Mk_INMDF}
The moments $M_q$ of the generalized distribution function given by \refeq{eq.fnM.Generalization} become
\bgeqa
\label{eq.fnM.Generalization.Mq}
\displaystyle
M_q &=& \sum_{k=0}^{N} \Bigg[ \Delta_k \sum_{p=0}^{n_k} \left(\begin{array}{c}n_k\\p\end{array}\right) \left(-b_k\right)^{(n_k-p)} 
J_{p+q}\left( \frac{1}{2e_k}, \frac{c_k}{e_k} \right) \Bigg],
\edeqa
where $\Delta_k = a_k \left(2\pi d_k^{m_k}\right)^{-1/2} \exp\left( - c_k^2 / ( 2e_k ) \right)$, $\left(\begin{array}{c}n_k\\p\end{array}\right)$ is the combination between the integers $p$ and $n_k$, and the functions $J_k(a,b)$ are introduced in Ref.~\citep{Izacard_2016_Paper1} and have been analytically evaluated with the help of Mathematica and Ref.~\citep{GR_2007}. For the 1D INMDF given by \refeq{eq.fI}, the first fluid moments are
\bgeqa
\label{eq.fI.M0}
\displaystyle
M_0 &=& n, \\
\displaystyle
M_1 &=& n v + \Gamma, \\
\displaystyle
M_2 &=& n \left( T + v^2 \right) + 2 \Gamma c,  \\
\displaystyle
M_3 &=& n v \left( 3 T + v^2 \right) + 3 \Gamma \left( W + c^2 \right), \\
\displaystyle
M_4 &=& n \left( 3 T^2 + 6 T v^2 + v^4 \right) + 4 \Gamma c \left( 3 W + c^2 \right), \\
\displaystyle
M_5 &=& n v \left(15 T^2 + 10 T v^2 + v^4 \right) 
+ 5 \Gamma \left( 3 W^2 + 6 W c^2 + c^4 \right), \\
\label{eq.fI.M6}
\displaystyle
M_6 &=& n \left(15 T^3 + 45 T^2 v^2 + 15 T v^4 + v^6 \right) 
+ 6 \Gamma c \left( 15 W^2 + 10 W c^2 + c^4 \right).
\edeqa

\bibliographystyle{jpp}

\bibliography{jpp-references}

\end{document}